\magnification=\magstephalf    
\tolerance=10000
\hsize=15.3 truecm          
\vsize=22 truecm            
\newcount\initpage
\initpage=0
\overfullrule=0pt
\mathsurround=2pt       
\binoppenalty=10000     
\voffset= -1 truecm    
\baselineskip=15truept     
\long\def\centro#1{\hfill#1\hfill}
\def\vs{\vskip .3truecm}
\def\1{\vskip .3truecm}
\def\2{\vskip .6truecm}
\def\3{\vskip .9truecm}
\def\M{{\cal M}}
\def\R{{I\!\! R}}
\def\non{{\vskip .5truecm\noindent}}
\def\no{{\noindent}}

\def\ep{{\epsilon}}
\def\p{\partial}

\def\die{~}
\def\cvd{{\hbox{\enspace${\bf \square}$}} \smallskip}
\def\sqr#1#2{{\vcenter{\vbox{\hrule height .#2pt
\hbox{\vrule width .#2pt height#1pt \kern#1pt
\vrule width .#2pt}
\hrule height .#2pt}}}}
\def\square{\mathchoice\sqr54\sqr54\sqr{4.1}3\sqr{3.5}3}
\def\io{{\infty}}
\def\r{\mathop{{\rm I}\mskip-4.0mu{\rm R}}\nolimits}
\def\k{\mathop{{\rm I}\mskip-4.0mu{\rm K}}\nolimits}
\def\n{\mathop{{\rm I}\mskip-4.0mu{\rm N}}\nolimits}
\def\L{{\cal L}^{+,2}_{p,\gamma}(\Lambda)}
\def\Q{{\cal Q}^{+,2}_{p,\gamma}(\Lambda)}
\def\Om{{\Omega}^{+,2}_{p,\gamma}(\Lambda)}
\def\P{{\cal P}^{+,2}_{p,\gamma}(\Lambda)}
\def\N{I\!\!N}
\centerline{{\bf THE FERMAT PRINCIPLE IN GENERAL RELATIVITY}}
\centerline{{\bf AND APPLICATIONS}${}^*\!\!$}
{\footnote{}{\rm ${}^*$ The first two authors were partially sponsored by 
research funds M.U.R.S.T.\ (Italy), 
and the third author by CNPq (Brazil), Processo n.\ 301410/95-0.}}
\centerline{}
\centerline{{\bf F. Giannoni}}
\centerline{Dipartimento di Matematica e Fisica}
\centerline{ Universit\'a di Camerino}
\centerline{{\it e-mail:} \tt giannoni@campus.unicam.it}
\medskip
\centerline{\bf A.  Masiello}
\centerline{Dipartimento di Matematica}
\centerline{ Politecnico di Bari, Italy}
\centerline{{\it e-mail:} \tt masiello@pascal.dm.uniba.it}
\medskip
\centerline{{\bf P. Piccione}}
\centerline{Instituto  de  Matem\'atica  e Estat\'\i stica}
\centerline{Universidade de Sao Paulo, Brazil}
\centerline{{\it e-mail:} \tt piccione@ime.usp.br}
\centerline{}
\centerline{}

\centerline{\bf ABSTRACT}
\bigskip
\centro{\vbox{\hsize10truecm
In  this paper  we use a general version of Fermat's principle for light 
rays  in General Relativity and a curve shortening method to write
the Morse relations for light rays joining an event with a smooth timelike
curve in a Lorentzian manifold with boundary.
 As  a  physical  meaning,  one  can  apply  the Morse relations to
have a mathematical description of the {\it  gravitational lens
effect} in a very general context. 
}}
\vfil
\eject

\centerline{\bf INTRODUCTION}
\1
The Fermat's Principle in Classical Optics
states that the trajectory of a light ray
from a source $A$ to a target $B$ is such
that it is a minimizer, or better, a stationary
curve for the travel time among all the
paths joining the points $A$ and $B$.
This variational principle can be extended in
the context of General Relativity where the trajectory
of a light ray under the action of the gravitational
field in vacuum is given by a null geodesic in a Lorentzian
manifold modelling
the space--time generated by a gravitational mass distribution.

A  formulation of the Fermat's principle is given once the
following data are determined:
\smallskip

\item{1.} a set of {\sl trial curves\/} joining the light source
and the observer;
\smallskip
\item{2.} a {\sl functional\/} that associates to each trial curve
a real number, which has to be related to a measurement of the {\sl time\/} 
passed from the instant at which the photon departed from the
light source to the instant at which the photon arrives 
to the observer.
\smallskip
A mathematical proof of the Fermat's principle consists in proving
that the trajectory of a light rays is {\sl characterized\/}
as a stationary point of the time functional in the set of trial
curves. 
The geodesics in a semi-Riemannian manifold are characterized
as solutions of differential equations, and the {\sl local\/} theory
of the light rays can be developed in terms of systems of
differential equations in $I\!\!R^n$. However, the variational approach
has the advantage of providing techniques for proving {\sl global\/}
existence results, and also for producing several kinds of estimates
on the number of solutions, given in terms of the topology of the space
of trial curves. 
To this aim in this paper we prove the 
{\sl Morse Relations\/} for light rays, that will be presented
in details in Section~1. We now proceed to a general discussion of
the mathematical problem, its physical applications and
a presentation of the new results that will be proven in this paper.
We fix a Lorentzian manifold $({\cal M},g)$ that is the mathematical
model of our relativistic spacetime, and we assume
that $\cal M$ is endowed with a time orientation given by the
choice of a continuous timelike vector field $W$ on $\cal M$. 
Such assumption is indeed very mild; namely, given any Lorentzian
manifold, there exists always a two-fold covering $\widetilde{\cal M}$
of $\cal M$ that admits a time orientation
(cf. [19]), and clearly there is a two-to-one
correspondence between the geodesics in $\widetilde{\cal M}$ and those
on $\cal M$. If we want to study the light rays emitted by some source
at a given time in the past, represented by an event $p$ of $\cal M$,
and reaching an observer sometimes during its life, whose worldline
is given by a timelike curve $\gamma$ in $\cal M$, then we need to
determine all the lightlike future pointing  geodesics joining
$p$ and $\gamma$ in $\cal M$. 
We are assuming here that both the source and the
receivers are {\sl pointlike}, i.e. 
they have dimensions which are neglectible 
with respect to their distance; a variational principle for light rays
between a spatially extended source and a spatially extended receiver
may be found in [23].
In analogy with the principle in classical optics, the set of
trial curves is chosen to be the set of all possible future pointing
trajectories joining the source with the
observer, and that are run at the speed of light. This amounts 
to saying that a trial curve is a curve whose tangent vector is 
everywhere in the light cone, and it belongs to the same half light cone
as the vector field $W$. 
The choice of the {\sl regularity\/} to impose on the trial curves and,
most of all, the choice of the functional to be extremized are
rather delicate questions, which have deep consequences for the 
mathematical theory to be developed. 
The first relativistic formulation
of the principle, valid in the case of a {\sl static\/}
spacetime,  is due to Weyl (see [33]); the validity of
the general relativistic Fermat's principle was successively
extended to the case of {\sl stationary\/} spacetimes
by Levi--Civita (see [15]). For {\sl conformally stationary\/}
spacetimes, an alternative formulation of the principle is
given in [6].  
The first attempt to extend the Fermat's principle beyond the (conformally)
stationary case is due to Uhlenbeck (see [31]), who considered a Lorentzian
manifold diffeomorphic to a space-time splitting 
${\cal M}_0\times I\!\!R$ and a 
time dependent metric which is diagonal  with respect to this product.
The variational principle proven in [31] employs the time functional
given by the the projection onto the second factor calculated at the
final point of each trial curve. Such functional does not depend on the
parameterization of the trial curve as, for instance, the {\sl length\/}
functional for curves in a Riemannian manifold, 
and this lack of {\sl rigidity\/} 
makes it a  difficult task to obtain results of existence and multiplicity
of critical points. 
For this reason, in order to prove the classical
Morse relations the author employs an action 
functional whose Lagrangian function 
depends  quadratically  by  the  velocities. 
This kind of functional has a 
strict  relationship  with  the 
{\sl energy\/} functional for Riemannian geodesics, 
obtained  by  {\sl removing the square root\/} inside the integral
that defines the length.  
The same variational  principle  was  used  in [8] to 
obtain Morse relations for light 
rays  on  orthogonal  splitting  Lorentzian  manifolds,  using an infinite 
dimensional setting 
and covering some gaps that occur in [31]. 
Such a variational 
principle  was  extended in  [1] to 
{\sl stably causal\/} Lorentzian manifolds and 
applied in [9,10] 
to obtain multiplicity results and Morse Relations for light 
rays joining an event $p$ with a timelike curve $\gamma$ in the presence
of a smooth convex boundary.
A very general version of the principle, valid in all spacetimes,
was given recently by Kovner (see [14]), who introduced the so
called {\sl arrival time\/} functional with respect to the observer
$\gamma$, defined  on the space of {\sl piecewise smooth\/}
lightlike  curves  joining  $p$  and  $\gamma$. Such functional is given
by fixing any (future pointing) parameterization of $\gamma$, and assigning
to each trial curve the value of the the parameter of $\gamma$ at the
arrival point. Any two future pointing
parameterizations of $\gamma$ differ by an order preserving diffeomorphism
between two intervals of the real line; 
it is an easy observation that the stationary
points of the arrival time functional do not indeed depend on the choice
of the parameterization of $\gamma$.
A  rigorous mathematical proof of the Kovner's claim was given in [21].
However, the  proof in [21]  needs the assumption  that the critical points 
have nonzero derivative everywhere.
An alternative variational principle on the space of lightlike curves $z$
with    a   suitable   prescribed  parameterization  and  satisfying $\dot 
z(s)  \neq  0$   for  all  $s$,  can  be  found in [22]. 
However, using this approach it is not  
possible   to obtain Morse Relations, as 
it will be clear from the discussion presented in Appendix~B.
In reference [10]  the reader will find a more detailed presentation
of the different versions of the Relativistic Fermat  Principle  and  
some examples and applications to 
the multiple image effect (the so called "gravitational lens effect").
In a paper published in 1979, Walsh, Carlswell and Weymann 
(see [32]) discussed
the possibility that the double quasar 0957+561 
would be a good candidate for a
gravitational lens effect. 
Such a name refers to the phenomena occurring when a
multiple image of some stellar object is observed. 
The multiple image effects
are due to the deflection of the light in presence of a 
gravitational field. 
We refer e.g. to [27,28] for a detailed physical description of the 
gravitational lens effect and many physical examples.
The version of the relativistic Fermat principle introduced by Kovner
allows also to treat nonstationary situations such as a gravitational wave
sweeping over a gravitational lensing situation. More details can be found
in [5].

Some natural questions arise in the study of the gravitational
lensing effect; for instance, it is tried to understand
under which circumstances a multiple imaging of a distant
source can occur, and, in this case, how many images  of the source
can be seen.
In mathematical terms, these questions can be answered by giving
conditions on the topology and the metric of the spacetime that
guarantee a multiplicity of lightlike geodesics between $p$
and $\gamma$ lying inside an open set $\Lambda$, that represent
the region of the universe where to localize the description for any
gravitational lens.
As already mentioned, 
a technique for investigating these issues is provided by
the Morse Theory, 
which is a well established mathematical theory that relates
the critical points of a smooth functional with the topology of the underlying
space. 

The main purpose of this paper is to develop an infinite dimensional
Morse theory under {\sl minimal\/} assumptions on the global
structure of the spacetime and on the timelike curve $\gamma$.
This in particular allows us to extend the results in [6,8,10,31]
concerning Morse Relations.
Most of all, we want to push the results beyond the compactness
assumption of {\sl global hyperbolicity\/} made in [8,31];
we also generalize the results of [10] in the following directions:
\smallskip
\item{$\bullet$} we do not assume the {\sl stable causality\/} of
the Lorentzian manifold $({\cal M},g)$, that will only be assumed to
be time orientable;
\item{$\bullet$} we do not assume any regularity  for
the boundary $\partial\Lambda$ of the region $\Lambda$;
\item{$\bullet$} we do not assume that 
$\gamma$ is embedded as a {\sl closed\/}
subset of $\Lambda\subseteq\cal M$.
\smallskip
Observe in particular that the second generalization above
allows to extend the results also to light rays  moving on 
a region of the universe exterior to a static blackhole (see [13]).
For the functional framework, we will employ Kovner's arrival time
functional, denoted by $\tau$; observe  that the definition of
$\tau$ does not require the existence of a {\sl global time function\/}
on $\Lambda$, which was a crucial assumption in [9,10].

For a correct physical interpretation of our results, all the relevant
information about the light rays joining $p$ and $\gamma$ must be
encoded the open subset $\Lambda$. For this reason, if $\Lambda\ne\cal M$,
we assume the following {\sl convexity\/} property of
$\Lambda$:
\vfil
\eject
$$\displaylines{
\hbox{{\it 
every lightlike geodesic starting from any event in ${\Lambda}$}} \cr
\hfill
\hbox{{\it 
and moving outside ${\overline \Lambda}$ does not come back in ${\Lambda}$.}} 
\hfill\llap{($\ast$)}\cr}
$$
Note that assumption $(\ast)$ 
is not strictly necessary to develop our theory.
As a matter of facts, we will use a more general assumption:
the light convexity of the boundary of $\Lambda$ (cf.(3) in Sect. 2).
Observe also that in the Minkowski space time a set $\Lambda_{0}~\times~\r$
satisfies condition $(\ast)$ precisely when $\Lambda_{0}$ is convex. 
Other simple examples of spacetimes
satisfying $(\ast)$ are the regions outside the event 
horizon of the Schwarzschild and Reissner-Nordstr\"om spacetimes (see [16]).
Our Morse relations are given for future pointing lightlike geodesics; 
we remark here that there is also a time-reversed version of the
Fermat's principle. Namely, $p$  can be
interpreted as a pointlike receiver at a particular instant of time 
and $\gamma$  as the worldline of a pointlike light 
source, in which case one is interested in determining the
past pointing light rays from $p$ to $\gamma$.
Clearly, the results proven in the paper are still valid in the past pointing
case.  
From a mathematical point of view, the case of past pointing light rays
it is completely analogous, and it will not be treated explicitly
in this paper.
In the next section we will give a formal
statement of our results, and we will present arguments
to  show that our assumptions can not be weakened to obtain a 
Morse Theory.

The reader is referred to classical books as [2,13,19] 
for the main notions and
properties in Lorentzian geometry.
Finally, we remark that alternative approaches to the study of
the Morse theory for light rays are available; 
for instance, in references [24,25,26]
the author applies the Morse Theory in a time
independent {\sl quasi-Newtonian\/} setting.
\2

\centerline{\bf 1. STATEMENT OF THE RESULTS AND}
\centerline{\bf DISCUSSION ABOUT THE ASSUMPTIONS}
\1
Let $(\M,g)$ be a smooth Lorentz manifold, $\Lambda$ an open connected
subset of $\M$, $p \in \Lambda$,
$\gamma \colon ]\alpha,\beta[  \longrightarrow  \Lambda$  a smooth  timelike
curve such that $p\notin \gamma(]\alpha,\beta[)$.
Here and in the rest of the paper we will often  set 
$\langle{\cdot},{\cdot}\rangle \equiv g(z)[\cdot,\cdot]$.
We assume that $(\M,g)$ is time orientable. 
This means that there  exists a smooth vector field $W$ on $\M$ 
such that $\langle W(z),W(z) \rangle < 0$
for any $z \in \M$. 
With respect to the orientation $W$ we assume that 
$$
\gamma \hbox { {\it is future pointing},}
\leqno(1)
$$
namely
$$
\langle \dot\gamma(s),W(\gamma(s)) \rangle < 0 \quad 
\forall s \in ]\alpha,\beta[.
$$ 
Since we want to study future pointing light rays joining $p$ and
$\gamma$  in $\Lambda$, 
 we only shall consider past and future relatively to $\Lambda$. 
More precisely given two points $q_1$ and $q_2$ in $\Lambda$, 
we say that $q_2$ is in the future of 
$q_1$ (in symbols $q_2 \in J^{+}(q_{1},\Lambda))$ if 
there exists a piecewise smooth curve 
$y : [0,1] \longrightarrow \Lambda$ such that 
$\langle \dot y,\dot y \rangle \le 0$ (i.e. $y$ is a causal curve), 
$\langle \dot y,W(y) \rangle < 0$ 
(i.e. $y$ is future pointing), $y(0)=q_1, y(1)=q_2$. 
In general if $A \subset \Lambda$ the future of $A$ (in $\Lambda$) is the set 
$$
J^{+}(A,\Lambda)=\bigcup_{a \in A} J^{+}(a,\Lambda),
$$
while the past of $A$ (in $\Lambda$) is the set 
$$
J^{-}(A,\Lambda)= 
\Big\{q \in \Lambda: A \cap J^{+}(q,\Lambda) \not= \emptyset. \}
$$
To have future pointing lightlike curves joining $p$ and $\gamma$ 
in $\Lambda$ clearly we need the following assumptions:
$$
\hbox {{\it 
there exists }} q_{+} \in \gamma(]\alpha,\beta[) \cap J^{+}(p,\Lambda).
\leqno(2)
$$
Moreover a light-convexity assumption on the closure 
$\overline \Lambda$ of the open subset $\Lambda$ is needed:
$$\displaylines{
\hbox{ {\it $\overline \Lambda$  is light convex,
i.e. all the lightlike geodesics in $\Lambda\cup\partial\Lambda$}}\cr
\llap(3)\hfill
\hbox {{\it with endpoints in $\Lambda$ are enterely contained in }} 
{\Lambda}.
\hfill
\cr}
$$
\noindent
Here $\partial \Lambda$ is the topological boundary of $\Lambda$.
Finally, to be able to define the arrival time functional we need
$$
\hbox {{\it $\gamma:\,]\alpha,\beta\,[\; \mapsto\Lambda$ is injective. }}
\leqno(4)
$$
By (4), on the space of the curves joining $p$ and $\gamma$ on the interval
$[0,1]$, it is well defined the arrival time functional
$$
\tau(z)=\gamma^{-1}(z(1)).
\eqno(1.1)
$$
The following assumptions says that $\tau$ 
is bounded from below on the set of the future pointing lightlike curves 
joining $p$ and $\gamma$.
$$
\hbox{ {\it there exists }} 
q_{-} \in \gamma(]\alpha,\beta[) \setminus J^{+}(p,\Lambda).
\leqno(5)
$$
Since we do not require that $\partial\Lambda$ is smooth 
we are not able to use the same penalizing argument as in [9,10] 
to overcome the difficulties due to the presence of the boundary. 
To develop a Morse Theory for the arrival time functional we should need a
flow which is strictly decreasing far from the critical points.
Since the boundary is not smooth a convenient approach is a shortening method.
Assumption (3) is necessary because we need a flow that it is 
invariant with respect to the lightlike curves with image in $\Lambda$. 
Note that to develop a Morse Theory 
the presence of the boundary is a difficulty to bypass because we 
want to treat with "free" critical points lying in $\overline \Lambda$. 
Note also that, even if the boundary would be 
smooth, a shortening method seems technically more simple that the penalized
techniques used in [8,9]. 
Moreover, since $\tau$ is invariant by reparameterizations,
as well as the space of future pointing light-like curves joining 
$p$ and $\gamma$, a shortening approach seems to be a good help to 
overcome also this kind of difficulty.

To use the shortening method we need an assumption assuring the existence of
minimizers in $\Lambda$ between events and timelike curves. For this
reason we need also the following assumption:
$$\displaylines{
\llap(6)\hfill
\hbox {{\it there exists a smooth timelike vector field $W$ in $\M$}
}\hfill\cr
\hbox{ {\it having the following properties}:}\cr
\llap(a)
\hfill
\hbox{ {\it $\gamma$ is an integral curve of $W$
(namely $\dot\gamma=W(\gamma)$ for any $s\in ]\alpha,\beta[),$ }} \hfill\cr
\llap(b)\hfill
\hbox {{\it for any  $q\in J^{+}(p,\Lambda)
\cup J^{-}(\gamma(]\alpha,\beta[),\Lambda)$ }} \hfill\cr
\hbox {{\it if $\gamma_{q}$ is the maximal integral curve of 
$W$ such that ${\gamma}_{q}(0)=q,$ }} \cr
\hbox {{\it there is 
$\overline q \in [Im\gamma_{q} \cap \Lambda] \setminus J^{+}(p,\Lambda).$
}} \cr}
$$
\noindent
Here ${\rm Im}\gamma_{q}$ denotes the image of the curve $\gamma_{q}$.
Note that (6) is certainly satisfied if $\Lambda$ is invariant with
respect to the flow of $W$.

Morse Theory gives an algebraic relation (in terms of formal series) between
the critical points of a suitable functional (in our case the arrival time
functional) and the topology of the space where the functional is defined.
At this point
there are two chances: 
to introduce a Sobolev space of lightlike curves or to use broken
lightlike geodesics as space of trial curves.
To state Morse relations we prefer here to use the second choice 
since it is not required the use of any auxiliary (Riemann) structure. 
Nevertheless in section 2
we shall give an infinite dimensional formulation of the 
Fermat Principle using Sobolev
spaces. 
Indeed,  even if we shall use a shortening procedure 
it is more convenient an infinite dimensional approach 
to study the arrival time functional
nearby its critical points. 
This is due to the fact that here it is hard to try to reduce
(as e.g. in [8,31])
the study the functional $\tau$ on a space of curves 
joining two given points.  
For this reason we are not able  to adapt to our case the Milnor 
finite dimensional approximation scheme (cf. [18]) nearby critical points.
 
Now set
$$\displaylines{
{\cal B}^+_{p,\gamma}({\Lambda})=
\Big\{z \colon [0,1] \longrightarrow {\Lambda} \colon~
z \hbox{ {\it is a $C^{2}$ piecewice curve such that} } \cr
z(0)=p,~z(1) \in \gamma(]\alpha,\beta[) 
\hbox{ {\it and,  on any interval $[a,b]\subset\,]\alpha,\beta\,[$
where $z$ is of class $C^{2}$,} } \cr
\hfill \hbox{ {\it $z$ is a constant or a future pointing
light-like geodesic} }\Big\}. \hfill\llap(1.2)\cr}
$$
We point out that a curve $z \in {\cal B}_{p,\gamma}^{+}(\Lambda)$ 
may be constant on some interval $[a,b] \subset [0,1]$ 
(and therefore $z_{|[a,b]}$ is not a light-like geodesic).
Nevertheless the topological structure of 
the problem is carried on by the space 
${\cal B}_{p,\gamma}^{+}(\Lambda)$, instead of the space 
$\hat {\cal B}_{p,\gamma}^{+}(\Lambda)$ of the broken lightlike 
geodesics (without subintervals where $z$ is constant).
A simple example in appendix B shows that Morse Relations can not be 
written using $\hat {\cal B}_{p,\gamma}^{+}(\Lambda)$.
In section 4 we shall prove the homotopy equivalence between 
${\cal B}^+_{p,\gamma}({\Lambda})$ (endowed with the uniform topology) 
and the Sobolev spaces of future pointing, lightlike, \
$H^{1,r}$-curves joining $p$ and $\gamma$ ($r\in [1,+\infty]$).
Using the space ${\cal B}^+_{p,\gamma}({\Lambda})$ we can state our last
assumptions. 
In section 2 we shall prove that it 
is equivalent to that one formulated in [9,10].
For any $c \in ]\alpha,\beta[$ (cf. (4))
we denote by
$\tau^c$ the $c$-sublevel of the functional $\tau$ in ${\cal
B}^+_{p,\gamma}({\Lambda})$:
$$
{\tau^{c}}~=~\Big\{z \in {\cal B}^+_{p,\gamma}({\Lambda}) 
\colon {\tau}(z)~\leq~c \Big\}
\eqno(1.3)
$$
\1

\non
{\bf Definition 1.1.} {\it
Fix $c>\alpha$. We say that ${\cal B}^+_{p,\gamma}({\Lambda})$ 
is $c-precompact$ if
any sequence $\{ z_{n}: n \in \n \} \subset {\tau}^{c}$ 
has a subsequence uniformly convergent
in $\overline\Lambda$, up to reparameterizations.
We say that $\tau$ is pseudo-coercive in 
${\cal B}^+_{p,\gamma}({\Lambda})$, if
${\cal B}^+_{p,\gamma}({\Lambda})$ is 
$c-precompact$ for any $c \in ]\alpha,\beta[.$}
\1

\noindent
Note that by assumption (5) there exists 
$\hat s > \alpha$ such that $\tau^c$ is the empty set for any 
$c \in ]\alpha,\hat s]$.)  
Whenever $\Lambda = {\cal M}$ pseudo--coercivity coincides with the global
hyperbolicity of the set of the events in the future of $p$ 
(cf. Proposition 2.13 and [9]).

Before stating our main result, we recall some definitions.
\1
\non
{\bf Definition 1.2.} {\it
Let $({\cal M},\langle\cdot,\cdot\rangle)$ be a Lorentzian manifold, 
and $z\colon[0,1] \longrightarrow {\cal M}$ be a geodesic.
A smooth vector field $\zeta$ along $z$ is called {\it Jacobi field} 
if it satisfies the equation}
$$
D^2_s \zeta + R(\zeta,\dot z)\dot z = 0~,
\eqno(1.4)
$$
{\it where $R$ is the curvature tensor of the metric
$\langle\cdot,\cdot\rangle$ (cf. {\rm [2]}).
A point $z(s)$, $s \in ]0,1]$ is said to be
{\it conjugate} to $z(0)$ along $z$
if there exists a nonvanishing Jacobi field $\zeta$ along $z_{|[0,s]}$ 
such that}
$$
\zeta(0) = \zeta(s) = 0~.
\eqno(1.5)
$$
{\it The {\it multiplicity} of the conjugate point $z(s)$ is the maximal
number of linearly independent Jacobi fields satisfying (1.5).}
\1
By (1.4) the set of the Jacobi fields is a vector space of dimension
$2{\rm dim} {\cal M}$. Hence  the multiplicity of a conjugate
point is finite and, by (1.5), is at most
${\rm dim}{\cal M}$ (actually it is at most
${\rm dim}{\cal M} - 1$ because $\zeta(s) = s \dot z(s)$ is a Jacobi field 
which is null only at $s = 0$).
\1
\non
{\bf Definition 1.3.}
{\it The index $\mu(z)$ is the number of conjugate points $z(s)$, 
$s \in ]0,1[$ to $z(0)$, counted with their multiplicity.}
\1
\noindent
It is well known that the index of a lightlike geodesic is finite (see
[2]).
\1
\non
{\bf Definition 1.4.}
{\it
Let $p$ be a point and $\gamma$ a timelike curve on a Lorentzian manifold
$({\cal M},g)$.
Then $p$ and $\gamma$ are said nonconjugate by 
lightlike geodesics if for any 
lightlike geodesic $z\colon [0,1] \longrightarrow {\cal M}$
joining $p$ and $\gamma$, $z(1)$ is nonconjugate to $p$ along $z$.}
\1
It is well known that such a condition is true except for  a  residual
set of pairs $(p,\gamma)$. 
For some results where $p$ and $\gamma$ are conjugate
see [7].
Let $X$ be a topological space and $\k$ a field. For any
$l \in \n$
let $H_l(X;\k)$ be the $l$-th homology group of  $X$
with  coefficients  in ${\k}$.
Since $\k$ is a field,
then $H_l(X;\k)$ is a vector space whose dimension
$\beta_l(X;\k)$ (eventually $+ \infty$) is called the
$l$-th {\it Betti  number  of $X$} (with coefficients in $\k$).
The {\it Poincar\'e polynomial}
${\cal P}(X;\k)$ is defined as the following formal series:
$$
{\cal P}(X;\k)(\kappa) =
\sum_{l \in \n} \beta_l(X;\k)\kappa^l~.
$$
Let 
${\cal G}^+_{p,\gamma}(\Lambda)$ be the set of the future pointing
lightlike geodesics joining $p$ and $\gamma$ 
and having image contained in $\Lambda$.
The main result of this paper is the following theorem.
\1
\non
{\bf Theorem 1.5.}
{\it Let
$\Lambda$, $p$, $\gamma$ satisfying {\rm (1)-(6)}.
Assume that the following assumptions hold true:}
\1
\centro{\vbox{\hsize14truecm
\item{${\rm L_1)}$}
{\it $p$ and $\gamma$ are nonconjugate;}
\item{${\rm L_2)}$}
{\it $\tau$ is pseudo-coercive on ${\cal B}^+_{p,\gamma}(\Lambda)$;}}}
\1
{\it Then for any field $\k$ there exists a formal series $S(\kappa)$
with coefficients in $\n \cup \{+\infty\}$, such that}
$$
\sum_{z \in
{\cal G}^+_{p,\gamma}(\Lambda)} {\kappa}^{\mu(z)} =
{\cal P}({\cal B}^{+}_{p,\gamma}(\Lambda);\k)(\kappa) + (1+\kappa)S(\kappa)~.
\eqno(1.6)
$$
\1
The  same  result  holds  for  the  lightlike  geodesics  joining   $p$
and $\gamma$ in the past of $p$, under an 
obvious modification of the assumptions.
\1
\non
{\bf Remark 1.6.}
Observe that the Betti numbers $\beta_l(X;\k)$ (and the coefficient
of the formal series $S(\kappa)$ in  (1.6))  depend  
in  a  substantial  way  on
the choice of the field $\k$. On the other hand, the left hand side of
the equality (1.6) does not depend on $\k$,  hence  one  can  obtain
more information on 
${\cal  G}^+_{p,\gamma}(\Lambda)$  by  letting  the  coefficient
field $\k$ arbitrary in (1.6).
For example, a result of Serre, where the choice of $\k$
is essential (cf. [27]), is used to prove part (b) of Theorem 1.11.
\1
\non
{\bf Remark 1.7.}
Note that assumption  ${\rm  L_2)}$   cannot   be   removed.
Indeed,  let  $({\cal  M}_0,\langle\cdot,\cdot\rangle_0)$  be  a
Riemannian manifold such that there exist two points
$p_1$, $p_2 \in {\cal M}_0$ which are not joined by  any   geodesic  for
the metric  $\langle\cdot,\cdot\rangle_0$.
Consider the (static)   Lorentzian manifold
$({\cal M}, \langle\cdot,\cdot\rangle)$, where
${\cal M} = {\cal M}_0 \times \r$ and
$\langle\cdot,\cdot\rangle$ is given by
$$
\langle\zeta,\zeta\rangle = \langle\xi,\xi\rangle_0 - \theta^2~,
$$
for any $z = (x,t) \in {\cal M}_0\times \r$ and
$\zeta = (\xi,\theta)  \in T_z{\cal M}$. Let $\Lambda = \M$. 
Consider the point  $p  =  (p_1,0)$  and  the  timelike  curve  $\gamma(s)  
= (p_2,s)$.
Clearly assumption (1)-(6) are satisfied, 
but not ${\rm  L_2)}$. Theorem  1.5  does
not hold for $p$ and $\gamma$, since there are no lightlike geodesics
joining $p$ and $\gamma$, while
${\cal P}({\cal B}^{+}_{p,\gamma}(\Lambda),\k)(\kappa) \not= 0$
for any field $\k$.
\1
\non
{\bf Remark 1.8}.\die
Let $c_l$ be the number of the future pointing 
lightlike geodesics joining $p$  and $\gamma$ having index $q$.
Then (1.6) can  be written in the following way:
$$
\sum_{l=0}^\infty c_l {\kappa}^l =
\sum_{l=0}^\infty\beta_l({\cal B)}_{p,\gamma}^{+}(\Lambda);\k){\kappa}^l +
(1+{\kappa})S(\kappa)
\die.
\eqno(1.7)
$$
{}From (1.7) we deduce that a certain number of  future pointing light
rays joining $p$ and
$\gamma$ are obtained according to the  topology  of
${\cal B}_{p,\gamma}^{+}(\Lambda)$.
In particular, setting $\kappa = 1$ in (1.7), 
we have the  following  estimate
on the number
${\rm card }({\cal G}^+_{p,\gamma}(\Lambda))$ of 
the light rays joining $p$ and $\gamma$:
$$
{\rm card }({\cal G}^+_{p,\gamma}(\Lambda)) =
\sum_{l=0}^\infty\beta_l({\cal B}_{p,\gamma}^{+}(\Lambda);\k) + 2S(1)\die .
\eqno(1.8)
$$
Since $S(1)$ is nonnegative we also  get the classical Morse inequalities
$$
c_l \geq \beta_l({\cal B}_{p,\gamma}^{+}(\Lambda);\k)\die ,
\quad\quad
\forall l \in {\bf N}\die .
\eqno(1.9)
$$
\1
An example of the influence of the topology of 
${\cal  B}^+_{p,\gamma}(\Lambda)$
on the number of future pointing, lightlike geodesics between $p$
and $\gamma$ is given by the next Theorem.
\1
\non
{\bf Theorem 1.9.}
{\it Under the assumptions of {\rm Theorem 1.5} we have:}
{\rm (a)}
{\it If ${\cal B}_{p,\gamma}^{+}(\Lambda)$ is contractible
the number ${\rm card }{\cal G}^{+}_{p,\gamma}(\Lambda)$ is infinite or odd;}
{\rm (b)}
{\it if ${\cal B}_{p,\gamma}^{+}(\Lambda)$ is not contractible, there  exist
at least two future pointing light rays joining $p$ and $\gamma$}.
(We recall  that  a  topological  space  is  said to be 
{\it  contractible} if  it is homotopically equivalent to a point.)
\1
Actually, the topology of ${\cal B}_{p,\gamma}^{+}(\Lambda)$ 
is in general not known for
arbitrary Lorentzian manifold.
More information can be obtained if its  topology can
be  related to  the  topology of the  manifold $\Lambda$.
Let $\Omega(\Lambda)$ be the {\it based  loop  space}  of  all
the  continuous  curves 
$z\colon [0,1]\longrightarrow\Lambda$ such that $z(0) = z(1) = \bar z$.
Since $\Lambda$ is connected, $\Omega(\Lambda)$ does not depend on $\bar
z$. 
We equip $\Omega(\Lambda)$  with  the {\it uniform topology}.
Since  the Poincar\'e polynomial is a homotopical  invariant,
we have the following result as an   immediate consequence of Theorem 1.5.
\1
\non
{\bf Theorem 1.10.}
{\it Besides the assumptions of {\rm Theorem 1.5}, assume also}
\smallskip
\noindent
${\rm L_3)}$
{\it
${\cal B}^+_{p,\gamma}(\Lambda)$ has  the  same  homotopy type  of  the  based
loop  space $\Omega(\Lambda)$.}
\smallskip
{\it Then for any field ${\cal K}$ there exists a formal series $S(\kappa)$
with coefficients in $\n\cup\{+\infty\}$, such that}
$$
\sum_{z \in {\cal G}_{p,\gamma}^+(\Lambda)} r^{\mu(z)} =
{\cal P}(\Omega(\Lambda),\k)(\kappa) + (1+\kappa)S(\kappa)~.
\eqno(1.11)
$$
\1
In appendix A we give a general condition assuring that 
${\rm L_3)}$ is satisfied. 
Thanks to the results proved in section 4 and appendix A 
we see that assumption  ${\rm  L_3)}$  is  certainly  satisfied  if
$(\Lambda,\langle\cdot,\cdot\rangle)$   is 
conformally stationary (for instance if 
$(\Lambda,\langle\cdot,\cdot\rangle)$ is a  Robertson-Walker 
space-time).  
\1
\non
{\bf Theorem 1.11.}
{\it Under the assumptions of Theorem 1.10 we have:}
\smallskip
{\rm (a)}
{\it If $\Lambda$ is contractible, then the number of the future pointing
lightlike geodesics joining $p$  with   
$\gamma$ and with image in $\Lambda$ is infinite or odd;}
\smallskip
{\rm (b)}
{\it if $\Lambda$ is noncontractible, then the number of the future pointing
lightlike geodesics joining $p$  with   
$\gamma$  and with image in $\Lambda$ is infinite.}
\1
Conditions assuring the finiteness of the images can be found in [12].
To write Morse Relations for light rays using the arrival time 
functional $\tau$ it does not seem that the pseudocoercivity assumption
can be weakened. 
Indeed it is necessary to have that the sublevels
$\tau^{c}$ of the arrival time functional are complete with respect to a 
suitable metric
and the Palais-Smale sequences are precompact 
(with respect to such a metric).  
Note that the sequences in Definition 1.1 are allowed to reach
$\partial\Lambda$. 
The light convexity of $\overline \Lambda$ will guarantee the 
existence of minimizers with image entirely included in $\Lambda$.

Morse Relations are proved regarding lightlike geodesics 
as critical points of the functional $\tau$. 
They are written using the geometric index $\mu$ instead of the Morse 
index thanks to Theorem 4.14. 
Its proof is based on a different approach to the Index Theorem 
for lightlike geodesics, with respect to that one of [2], where  
the Index Theorem is proved on a quotient space of the admissible
variations. 
For the proof of Theorem 4.14 we have to choose a suitable manifold 
where the critical points  of $\tau$ are lightlike geodesics.
\2

\centerline{\bf 2. MINIMIZERS FOR THE ARRIVAL TIME}
\centerline{\bf ON SOBOLEV CURVES SPACES}
\1

We begin the  section introducing the Sobolev spaces
$H^{1,r}([0,1],\Lambda)$ with $r\in [1,+\io]$. This can be rapidly done in
the following way.

\no 
Let $W$ be the smooth timelike vector field on $\M$ 
whose existence is assumed in (1). 
The manifold
$\M$ can be equipped by a natural Riemannian structure setting
$$
\langle \zeta,\zeta\rangle_R=\langle \zeta,\zeta\rangle-{2\langle
W(z),\zeta\rangle^2\over \langle W(z),W(z)\rangle.}
\eqno(2.1)
$$
The Riemannian
metric (2.1) can be used to introduce a Riemann distance on $\Lambda$ that
we shall denote by $d_R$. Such a distance allows us introduce the space of
the absolutely continuous curves between [0,1] and $\Lambda$ (denoted by
$AC([0,1],\Lambda)$). 
Finally for any $r\in [1,+\io[$ we set
$$
H^{1,r}([0,1],\Lambda)=\Big\{z\in AC([0,1],\Lambda):\int_0^1(\langle\dot
z,\dot z\rangle_R)^{r/2} ds<+\io\Big\}
$$
while
$$
H^{1,\io}([0,1],\Lambda)=\Big\{z\in AC([0,1],\Lambda):
\sup\big\{\langle\dot
z(s),\dot z(s)\rangle_R: s\in [0,1]\big\}<+\io\Big\}.
$$
Using local coordinates and the Palais definition of Sobolev manifolds 
(cf. [20]) we see that the spaces defined above 
do not depend on the choice of $W$.
Moreover we set
$$
\Omega^{1,r}_{p,\gamma}(\Lambda)=\Big\{z\in
H^{1,r}([0,1],\Lambda):z(0)=p,\ \
z(1)\in\gamma(\,]\alpha,\beta[\,)\Big\}
\eqno(2.2)
$$
For any
absolutely continuous curve $z$ we can extend the classical definition
of causal curve saying that $z$ is a causal curve
if $\langle \dot z(s),\dot z(s) \rangle \leq 0$ almost everywhere (a.e.). 
Moreover we say that a causal curve is future pointing if
$$
\langle \dot z(s),W(z(s))\rangle < 0\ \ \hbox{{\it for almost every $s$ such
that}}\ \ \dot z(s)\ne 0.
$$
It is possible to prove that the above notions are equivalent to that ones
given in [13] for
continuous curves, whenever we deal with absolutely continuous curves. 
This can be done using Proposition 2.2.
To develop a Morse theory
for light rays the following spaces will be also used:
$$
{\cal L}^{+,r}_{p,\gamma}(\Lambda)=
\Big\{z\in\Omega^{1,r}_{p,\gamma}([0,1],\Lambda):
\langle\dot z,\dot z\rangle=0 \hbox{ a.e. and }
z \hbox{ is future pointing} \Big\}.
\eqno(2.3)
$$
More in general for any event $p_*$ and any future pointing, 
injective, timelike curve 
$\gamma_* : ]\alpha_*,\beta_*[ \longrightarrow \Lambda$
we shall use the following notation:
$$\displaylines{
{\cal L}^{+,r}_{p_*,\gamma_*}([a,b],\Lambda) = 
\Big\{z \in H^{1,r}([a,b],\Lambda): \cr
\langle\dot z,\dot z\rangle=0 \hbox{ a.e., }
z \hbox{ {\it is future pointing, }}  
z(a)=p_* \hbox{ and } z(b) \in \gamma_*(]\alpha_*,\beta_*[\Big\}. \cr}
$$ 
We shall denote by the same symbol $\tau$ the functional defined by 
$\gamma_{*}^{-1}(z(b))$ on the space
${\cal L}^{+,r}_{p_*,\gamma_*}([a,b],\Lambda)$.
\1
\noindent
{\bf Remark 2.1.} Note that the above spaces are not smooth manifold: a
tangent spaces is not well defined on the curves 
$z$ such that $\dot z(s)=0$ on a
subset of $[0,1]$ having positive Lebesgue measure.
\no 
For this reason if we want to deal with smooth manifolds we need to use
an approximation of 
${\cal L}^{+,r}_{p,\gamma}(\Lambda)$ by suitable smooth manifolds
and to study apriori estimates for the limit process (cf. [9,10]). 
In this paper we shall work directly on 
${\cal L}^{+,r}_{p,\gamma}(\Lambda)$ showing
first that assumptions (1)--(6) and pseudocoercivity allow 
to find smooth minimizers
in $\Lambda$ which are lightlike geodesics.
\no 
This is a further motivation to choose a shortening procedure for the
arrival time functional, 
since it permits to bypass the nonsmoothness of the spaces defined by
(2.3). 
It will be possible to write Morse Relations using the Poincar\'e  
polynomial of 
${\cal B}^{+}_{p,\gamma}(\Lambda)$ because we shall prove in Sect. 3 
that it is homotopically equivalent to 
${\cal L}^{+,r}_{p,\gamma}(\Lambda)$.  
\1
\noindent
The next result is a local version of the relativistic Fermat 
Principle and it is the first step for the shortening procedure.
\non
{\bf Proposition 2.2.} {\it Let $q \in {\cal M}$. 
Then there exists $\rho(q) > 0$ having the following property. 
For any $\gamma_*$ timelike injective curve and integral curve of $Y$ 
such that:
\item{$\bullet$} $0 < d_R(q,Im\gamma_*) \leq \rho(q),$
\item{$\bullet$} $Im\gamma_* \cap J^{+}(q,\M) \not= \emptyset,$
\item{$\bullet$} $Im\gamma_* \setminus J^{+}(q,\M) \not= \emptyset,$
\noindent
there exists a unique future pointing light-like geodesic joining $q$ 
and $\gamma_*$ and minimizing the arrival time on 
${\cal L}^{+,r}_{q,\gamma_*}(\M).$}
\1

The proof can be e.g. obtained as a limit process of 
timelike problems using the results of [11]. 
Anyway here we shall give a variational 
proof working directly in the lightlike case. 
The proof is quite different from the classical geometrical one (cf. [13]).
The following Remarks will be used for the proof of Proposition 2.2 and 
other results in the present paper.
\1
\noindent
{\bf Remark 2.3.}  For any $z\in\Lambda$ there exists a neighborhood
${\cal U}_z$ of $z$ and a coordinate system $\varphi=(x_1,\ldots
x_{N-1},t)$ $(N={\rm dim}\, \Lambda)$ on 
${\cal U}_z$ such that $W={\p\over\p t}$
and ${\cal U}_z=\Sigma\times\, ]a,b[$ where $\Sigma$ is a spacelike
hypersurface parameterized by $x_1,\ldots,x_{N-1}$.
Moreover, in the coordinate $x=(x_1,\ldots,x_{N-1})$ and $t\in\,]a,b[$ the
metric $g$ is given by
$$
g(x,t)[(\xi,\theta),(\xi,\theta)]=\langle\alpha(x,t)\xi,\xi\rangle_0
+2\langle\delta(x,t)),\xi\rangle_0\theta-\beta(x,t)\theta^2 ,
$$
where $(\xi,\theta)\in T_{(x,{{a+b}\over {2}})}\Sigma\times\r$, 
$\langle \cdot,\cdot \rangle_0$ is the
restriction of $g~to~\Sigma,\ \alpha$ 
is a smooth, symmetric
positive definite operator, $\delta$ is a smooth vector field on
$\Sigma$ and $\beta$ is a smooth positive real function.
Note that $\langle \cdot,\cdot \rangle_0$ is a Riemannian metric on $\Sigma$.
Indeed it is sufficient to choose $\Sigma=\{(x,t):\ t={a+b\over
2}\}$, $\alpha(x,t)$, $\delta(x,t)$ and $\beta(x,t)$ such that
$$\displaylines{
\langle\alpha(x,t)\xi_1,\xi_2\rangle_{0}=g(x,t)[\xi_1,\xi_2] \hbox { for any }
\xi_1,\xi_2\in T_{(x,{{a+b}\over {2}})}{\Sigma}, \cr
\delta(x,t)=[G_0(x,t)]^{-1}(\bar\delta(x,t)),\
\bar\delta(x,t)=\sum\limits_{i=1}^{N-1} g_{iN}(x,t){\p\over\p x_i}, \cr
G_0(x,t)=(g_{i,j})_{i,j=1,...,N-1}, \cr}
$$
and $\beta(x,t)=-g(x,t)[W,W]$. 
\non
{\bf Remark 2.4.} We will assume henceforth that $W$ is {\sl renormalized}
in such a way that: 
$$\langle W(z),W(z)\rangle=-1\ \ {\rm for\ any}\ \ z\in\M.$$
In particular, $\langle \dot \gamma, \dot \gamma \rangle \equiv -1$,
and so the parameter of $\gamma$ can be interpreted as proper time.
Therefore, in the coordinate systems $(x_1,\ldots, x_{N-1},t)$ with
${\p\over\p t}=W$ (cf. Remark 2.3), 
$z=(x,t)\in{\cal L}^{+,r}_{p,\gamma}(\Lambda)$ if
and only if
$$
\dot t=\langle \delta,\dot x\rangle_0 + 
\sqrt{
\langle \delta,\dot x\rangle_0^2+
\langle \alpha\dot x,\dot x\rangle_0 
}
\eqno(2.4)$$
because
$$\beta(x,t)=-\langle W(z),W(z)\rangle \equiv 1.$$
Moreover, in such a coordinate system, 
any integral curve of $W$ can be written as
$$s\mapsto (\bar x,s),$$
for some $\bar x \in \Sigma$.
\1
\noindent
In order to prove Proposition 2.2, the following preliminary results are
needed. 
\1
\noindent
{\bf Lemma 2.5.}
{\it Under the assumptions of Proposition 2.2, if $\rho (q)$ is
sufficiently small, there exists a minimizer of $\tau$ on
${\cal L}^{+,1}_{q,\gamma_\ast}({\cal M})$ having $H^{1,\infty}$ regularity.}
\1
\noindent
{\bf Proof.} 
By Remarks 2.3--2.4 and the assumptions of Proposition (2.2), 
if $\rho(q)$ is sufficiently small, we can consider a sufficiently small 
neighborhood ${\cal U}$ 
of $q$ such that 
${\cal U} = V\times I$, where $V$ is an open neighborhood contained in
$\Sigma$, $I= ]-\lambda_0,\lambda_0[$, $q = (q_0,0) \in V\times I$, 
 the curve $\gamma_\ast(t) = (q_\ast,t)$ is defined in 
$]-\lambda_0,\lambda_0[$
and ${\rm d}_{\rm R}(q_0,q_\ast) \to 0$ whenever 
${\rm d}_{\rm R}(q,{\rm Im}\gamma_\ast) \to 0$. 
If 
$z \in {\cal L}_{p,\gamma}^+$ and takes its values in ${\cal U}$, 
then $z(s) = (x(s),t(s))$, $x(0) = q_0$, $x(1) = q_\ast$ and $t(s)$
satisfies 
the Cauchy problem
$$\cases{
\dot t =
\langle\delta(x,t),\dot x\rangle_{0} +
\sqrt{\langle\alpha(x,t)\dot x,\dot x\rangle_{0} 
+ \langle\delta(x,t),\dot x\rangle_{0}^{2}}
\cr
t(0) = 0\cr}\eqno(2.5)
$$
Moreover
$$
\tau(z) = 
t_x(1) = 
\int_0^1 \langle\delta(x,t_x),\dot x\rangle_{0}{\rm ds}
+
\int_0^1 
\sqrt{\langle\alpha(x,t_x)\dot x,\dot x\rangle_{0}
+ \langle\delta(x,t_x),\dot x\rangle_{0}^{2}}{\rm ds},
$$
where $t_x$ is the solution of the Cauchy problem above.
Using a minimal Riemannian geodesic between $q_0$ and $q_\ast$ in $V$ shows
that 
$$
\inf_{{\cal L}^{+}_{q,\gamma_\ast}} \tau \to 0  \quad 
\hbox{ as }
{\rm d}_{\rm R}(q,\gamma_\ast) \to 0.
$$
We show now the existence of a minimizer for $\tau$.
Consider a minimizing sequence $(z_m)$ for 
$\tau$ and set $z_m = (x_m,t_m)$. 
The coercivity of $\alpha$ shows that the Riemann length 
of $x_m$ is bounded. 
Then by (2.5) we deduce that $\dot t_m$ is bounded in $L^1$.

Let $\varphi_m$
be the solution of
$$\cases{
\dot \varphi_m = 
{{1} \over 
\displaystyle{
{-\langle \dot z_m(\varphi_m),W(z_m(\varphi_m))\rangle +1}}
} \cr
\varphi_m(0)=0 \cr}
$$
and set
$$
\psi_m(s) = \varphi_m(\lambda_m s),
$$
where $\lambda_m$ is such that $\varphi_m(\lambda_m) = 1$. 
Such a number exists since $\dot \varphi_m > 0$ 
and we can assume that $z_m$ is in $L^\io$ 
(choosing a minimizing sequence of  curves of class $C^1$), 
so that $\varphi_m$ is far from 0.

Then 
$$
\dot \psi_m = \lambda_m \dot \varphi_m(\lambda_{m}s) = 
{\lambda_{m} \over  
{-\langle \dot z_m(\psi_m),W(z_m(\psi_m))\rangle +1}}
$$
hence
$$
\lambda_m = 
\dot\psi_m\left[-\langle z_m(\psi_m),W(z_m(\psi_m))\rangle +1\right]
$$
and integrating over $[a,b]$ gives
$$
\int_a^b
\dot\psi_m\left[-\langle z_m(\psi_m),W(z_m(\psi_m))\rangle +1\right]
{\rm d}s = \lambda_m(b-a).
$$
Since 
$$
\int_a^b |\dot z_m|\;{\rm d}s \quad
\hbox{ is uniformly bounded},
$$
there exists a positive constant $M$ independent
on $m \in \N$ such that
$$
\Big\vert\int_a^b 
\langle \dot z_m,W(z_m)\rangle\;{\rm d}s\; \Big\vert
\leq M .
$$
Therefore the sequence $(\lambda_m)_{m\in\N}$ is bounded.
Now set $y_m = z_m(\psi_m)$. Then 
$$
\langle \dot y_m,W(y_m)\rangle = 
\dot\psi_m \langle \dot z_m,W(z_m)\rangle = 
-{\lambda_m \over
{-\langle \dot z_m, W(z_m)\rangle + 1}}
\langle \dot z_m,W(z_m)\rangle,
$$
and we deduce the existence of a positive constant $c_1$  such that
$$
|\langle \dot y_m,W(y_m)\rangle|\leq c_1,
$$
from which we deduce the existence of a positive constant $c_2$ such that
$$
\|\dot y_m\|_{L^\infty} \leq c_2.
$$
So we have a sequence of curves $z_m = (x_m,t_m)$ such that, up to a
reparameterization (and replacing $y_m$ with $z_m$), 
$$\displaylines{
(x_m,t_m) \hbox{ {\it uniformly converges to a curve }} z = (x,t);\cr
\hbox{ {\it the sequence $(\dot x_m)$ is equibounded in $L^\infty$;} }\cr
\hbox{ {\it the sequence $(x_m)$ weakly converges to $\dot x$ in 
$L^2$;} }\cr}
$$
In particular the sequence $(\dot x_m)$ weakly converges to $\dot x$ in 
$L^1$ and then by well known properties of the weak convergence (cf. [3]
for instance), 
$$\displaylines{
\int_0^1
\sqrt{
\langle\alpha(x,t)\dot x,\dot x\rangle_{0} 
+
\langle\delta(x,t)\dot x,\dot x\rangle_{0}^{2}}{\rm d}s \leq \cr
\liminf_{m\to\infty}
\int_0^1
\sqrt{
\langle\alpha(x,t)\dot x_m,\dot x_m\rangle_{0} 
+
\langle\delta(x,t)\dot x_m,\dot x_m\rangle_{0}^{2}}{\rm d}s.\cr}
$$
Moreover, we clearly have
$$\displaylines{
\left|
\int_a^b
\left(
\sqrt{
\langle\alpha(x,t)\dot x_m,\dot x_m\rangle_{0} 
+
\langle\delta(x,t),\dot x_m\rangle_{0}^{2}}
-
\sqrt{
\langle\alpha(x_m,t_m)\dot x_m,\dot x_m\rangle_{0} 
+
\langle\delta(x_m,t_m),\dot x_m\rangle_{0}^{2}}
\right)
{\rm d}s
\right| \leq \cr
\int_a^b
\left(
|\langle(\alpha(x_m,t_m) - \alpha(x,t)) \dot x_m,\dot x_m\rangle_{0} |
+
|\langle \delta(x_m,t_m) + \delta(x,t),\dot x_m\rangle_{0}
\langle \delta(x_m,t_m) - \delta(x,t),\dot x_m\rangle_{0}|\right)^{1/2}
{\rm d}s \cr
\longmapsto 0 \hbox{ as} ~ m \to +\infty.\cr}
$$
Finally, by the uniform convergence of $(x_m,t_m)$ to $(x,t)$, 
$$
\int_a^b 
\langle\delta(x_m,t_m),\dot x_m\rangle_{0}
\to 
\int_a^b 
\langle\delta(x,t),\dot x\rangle_{0}.
$$
It follows that for any $s_1 < s_2$, 
$$\displaylines{
\liminf_{m\to\infty}  t_m(s_2)- t_m(s_1) = \cr
\liminf_{m\to\infty}\left(
\int_{s_1}^{s_2}
\langle\delta(x_m,t_m),\dot x_m\rangle_{0} {\rm d}s 
+
\int_{s_1}^{s_2}
\sqrt{
\langle\alpha(x_m,t_m)\dot x_m,\dot x_m\rangle_{0} 
+
\langle\delta(x_m,t_m),\dot x_m\rangle_{0}^{2}}{\rm d}s\right) \cr
\geq
\left(
\int_{s_1}^{s_2}
\langle\delta(x,t),\dot x\rangle_{0}{\rm d}s 
+
\int_{s_1}^{s_2}
\sqrt{
\langle\alpha(x,t)\dot x,\dot x\rangle_{0} 
+
\langle\delta(x,t),\dot x\rangle_{0}^{2}}{\rm d}s\right).\cr}
$$
So we have obtained:
$$\cases{\displaystyle
\dot t \geq \langle\delta(x,t),\dot x\rangle_{0} 
+ 
\sqrt{
\langle\alpha(x,t)\dot x,\dot x\rangle_{0} 
+
\langle\delta(x,t),\dot x\rangle_0^{2}}\cr \cr
\displaystyle
t(0) = 0;\cr}
$$
Let $t_x$ be the solution of the Cauchy problem assigned above relatively
to the curve $x$. 
Comparison theorems for ordinary differential equations
show that 
$t_x\leq t$. 
Hence, $(x,t_x)$ is a minimizer for $\tau$.

Finally, our (reparameterized) minimizing sequence $z_m$ satisfies:
$$
\cases{ 
\langle \dot z_m,W(z_m) \rangle\  \hbox{\it  is bounded and}\cr \cr
\dot z_m\  \hbox{\it is weakly convergent to } \dot z \hbox{ in } L^2. \cr} 
$$
Therefore 
$\langle \dot z,W(z) \rangle$ is bounded 
and the minimizer $z$ is therefore in $H^{1,\io}$.
\cvd

\2
\noindent
{\bf Lemma 2.6.}
{\it 
Assume $q \notin \gamma_{\ast}(]\alpha,\beta[)$. 
Let $z$ be a minimizer of $\tau$ 
on ${\cal L}^{+,1}_{q,\gamma_\ast}({\cal M})$ 
such that $z \in H^{1,\io}([0,1],\M)$.
Then there exists a curve $y\in {\cal L}^{+,\infty}_{q,\gamma_\ast}$  
such that $y$ minimizes $\tau$ and}
$$
\inf\Big\{\|\dot y(s)\|_{\rm R} :s \in [0,1]\setminus N\Big\} > 0,
\eqno(2.6)
$$
{\it where $N$ is a subset of $[0,1]$ having null Lebesgue measure.
Moreover,}
$$
y([0,1]) = z([0,1]).
$$
\1
\noindent
{\bf Proof.}
For any $\delta > 0$, let$\varphi_\delta$ 
be the solution of the Cauchy problem
$$\cases{
\dot \varphi_\delta = 
{{-1}\over
\displaystyle{
{-\langle\dot z(\varphi_\delta), W(z(\varphi_\delta))\rangle 
+ \delta}}}
\cr
\varphi_\delta(0) = 0.\cr}
$$
Moreover, put $\psi_\delta(s) = \varphi_\delta(\lambda_\delta s)$, where
$\lambda_\delta$ is such that $\varphi_\delta(\lambda_\delta) = 1$. 
Such a number exists since $\dot\varphi_\delta > 0$ and it is far from zero because 
$\langle\dot z, W(z)\rangle$ is in $L^{\io}$.

Now 
$$
\dot\psi_\delta = 
{{\lambda_\delta}\over
{-\langle\dot z(\psi_\delta), W(z(\psi_\delta))\rangle 
+ \delta}}.
$$
Then, setting $y= z(\psi_\delta)$, we have:
$$
\langle\dot y,W(y)\rangle = 
\dot\psi\langle\dot z,W\rangle = 
{{\lambda_\delta}\over
{-\langle\dot z(\psi_\delta), W(z(\psi_\delta))\rangle 
+ \delta}}\langle\dot z,W\rangle,
\eqno(2.7)$$
while 
$$
\lambda_\delta = 
\left(-\langle\dot z(\psi_\delta),W(z(\psi_\delta))\rangle +
\delta\right)\dot\psi_\delta
= \int_0^1
\left(-\langle\dot z(\psi_\delta),W(z(\psi_\delta))\rangle +
\delta\right)\dot\psi_\delta {\rm  d}s \leq c_0
$$
where $c_0$ is a positive constant, because $\dot z$ is in $L^\io$.

Note that 
$$
\lambda_{\delta} = \int_{0}^{1} -\langle \dot z,W(z) \rangle{\rm d}s + \delta
\longrightarrow
\int_{0}^{1} -\langle \dot z,W(z) \rangle{\rm d}s \equiv \lambda_{0} 
\quad\quad \hbox{\it  as } 
\delta \to 0.
$$
Since $y_\delta = z(\psi_\delta)$ is equibounded in 
$L^\io$ (because $\dot \psi_\delta$ is equibounded and 
$\psi_\delta(0)=0$ we have (unless to consider $\delta_m \rightarrow 0$)
the existence of $y_0$ such that:
$$
\hbox{ 
$y_{\delta} \rightarrow y_0$
uniformly and 
$\dot y_{\delta} \rightarrow \dot y_{0}$   weakly in $L^2$. }
$$
Setting $y_0 = (x_0,t_0)$, arguing as in the proof of Lemma 2.5
we see that $t_0 = t_{\ast}$ where $t_{\ast}$ is the solution of (2.5)
with $x$ replaced by $x_0$. 
Then $\langle \dot y_{0}, \dot y_{0} \rangle = 0$ almost everywhere.
Moreover, since $\dot y_{\delta} \rightarrow \dot y_{0}$ 
weakly in $L^2$ we have
$$
\langle \dot y_{\delta},W(y_{\delta} \rangle \to 
\langle \dot y_{0},W(y_{\delta}) \rangle 
\quad\quad \hbox{ {\it  weakly in  $L^2$} },
$$
while
$$
-\langle \dot y_{\delta},W(y_{\delta} \rangle \geq 0 
\quad \quad\hbox{\it  a.e.}
$$
and, by (2.7)
$$
-\langle \dot y_{\delta},W(y_{\delta} \rangle \leq 
\lambda_{\delta} \quad \quad \hbox{\it  a.e.}.
$$
Then
$$
0 \leq -\langle \dot y_{0},W(y_{0} 
\rangle \leq \lambda_{0}
\quad
\quad \hbox{\it  a.e.}
$$
where $\lambda_0 = lim_{\delta \rightarrow 0}{\lambda_{\delta}}$. But 
$\lambda_{0} = -\int_{0}^{1} \langle \dot y_{0},W(y_{0}) \rangle$, hence
$\langle \dot y_{0},W(y_{0})\rangle = \lambda_{0}$ a.e. 
This implies that $\lambda_0 \not= 0$
(because $q \notin \gamma_{\ast}(]\alpha,\beta[)$) and 
therefore (2.6) follows with $y = y_0$.
Finally, by the definition of $y_\delta$ and the 
continuity of $y$ we deduce that $y$ and $z$
have the same image.
\cvd
\1
\noindent
{\bf Lemma 2.7.} 
{\it 
Let $z$ be a curve in ${\cal L}^{+,\io}_{p,\gamma}$ satisfying (2.6).
Then there exists a neighborhood 
$\cal V$ of $z$ in $H^{1,\io}([0,1],\Lambda)$
such that
${\cal V}  \cap {\cal L}_{p,\gamma}^{+,\infty}$ is a 
$C^1$--manifold and for
any $z\in \cal V$ its tangent space is given by}
$$\displaylines{
T_z({\cal L}^{+,\io}_{p,\gamma})=\Big\{\zeta\in
H^{1,\io}([0,1],T\Lambda):\zeta(s)\in 
T_{z(s)}{\Lambda}~for~any~s,\cr\qquad\qquad\quad
\zeta(0)=0,\ 
\zeta(1)\parallel\dot\gamma(\tau(z)),\ \langle D_{s}\zeta,\dot z\rangle=0\ \ 
a.e.\ \Big\}.\cr}$$
{\it Here $T\Lambda$ denotes the tangent bundle of 
$\Lambda$ and $D_s$ the covariant
derivative along $z$.}
\1
\noindent
{\bf Proof.} 
Consider the map 
$\psi:\Omega^{1,\io}_{p,\gamma}(\Lambda)\to L^\io([0,1],\r)$ such
that
$$\psi(z)=\sqrt 2 \langle \dot z,W(z)\rangle+\sqrt{\langle \dot z,\dot
z\rangle +2\langle \dot z,W(z)\rangle^2}=\sqrt 2\langle \dot
z,W(z)\rangle+\sqrt{\langle\dot z,\dot z\rangle_R}.$$
Note that $\psi^{-1}(0)={\cal
L}^{+,\io}_{p,\gamma}$. The set $\Omega^{1,\io}_{p,\gamma}(\Lambda)$ is a
manifold and, for any $z\in\Omega^{1,\io}_{p,\gamma}$ its tangent space is
$$\displaylines{
T_z\Omega^{1,\io}_{p,\gamma}=\Big\{\zeta\in H^{1,\io}([0,1],T\Lambda):
\zeta(s)\in
T_{z(s)}\Lambda\ \
{\rm for\  any}\ \ s\in [0,1],\cr
\zeta(0)=0,\ \ \zeta(1)\parallel\dot\gamma(\tau(z))\Big\}.}$$
By the above formula we immediately deduce that $\psi$ is of class $C^1$ 
in a neighborhood of $z$.
We claim that, $\forall z\in {\cal V} \cap
{\cal L}^{+,\io}_{p,\gamma}$, the differential $d\psi(z)[\cdot]$
is surjective.

Indeed let $U_z$ be the parallel transport of $\dot\gamma(\tau(z))$ along
$z$, i.e. the solution of the Cauchy problem
$$\cases{ D_{\dot z}U_z=0\cr U_z(1)=\dot\gamma(\tau(z))}$$
Then for any $\varphi\in L^\io([0,1],\r)$ it is easy to show the existence
of $\lambda\in H^{1,\io}([0,1],\r)$ such that $\lambda(0)=0$, 
$$d\psi(z)[\lambda U_z]=\varphi, $$
and the kernel of ${\rm d}\psi(z)$ splits.
Then $V\cap\psi^{-1}(0)$ is a manifold whose tangent space at $z$ is the
kernel of $d\psi(z)$ in $T_z\Omega^{1,\io}_{p,\gamma}$.\cvd

\non
{\bf Remark 2.8.} 
The functional $\tau$ is differentiable on $\Omega^{1,\io}_{p,\gamma}$. 
Since $\tau$ is defined by setting $\gamma(\tau(z)) = z(1)$, its differential 
satisfies
$$\dot\gamma(\tau(z))d\tau(z)[\zeta]=\zeta(1)$$
and therefore 
$$d\tau(z)[\zeta] = 0,$$
if and only if 
$$\zeta(1) = 0 \hbox{ for any } \zeta \in 
T_z({\cal L}^{+,\io}_{p,\gamma}).$$
Now let $V$ be a smooth vector field along 
$z$ such that $V(0)=V(1)=0$ and $U_z$ 
the parallel transport of $\gamma(\tau(z))$ along $z$. 
If $z$ satisfies (2.6) put 
$$\lambda(s)=\int_{0}^{s} \langle D_{r}V,
{
{\dot z} \over {\langle U_z,\dot z \rangle}
}
\rangle dr.$$
Then $V - \lambda U_z \in 
T_z({\cal L}^{+,\io}_{p,\gamma})$
and, if $z$ is a critical point of $\tau$ on
${\cal L}^{+,\io}_{p,\gamma}$ it is 
$$\tau(z)[V - \lambda U_z ]=0 \hbox{ for any } V$$
and therefore 
$$0 = \lambda(1)=\int_{0}^{1} \langle D_{s}V,
{
{\dot z} \over {\langle U_z,\dot z \rangle}
}
\rangle ds \hbox{ for any } V.$$
Conversely if $z$ satisfies the above condition, 
$\zeta (1) = 0$ for any $V$, hence $z$ is a critical point of $\tau$.   

\1
The following Theorem is the Relativistic Fermat principle proved 
in $H^{1,\infty}$ (cf. also [21]).
\2
\noindent
{\bf Theorem 2.9 (Fermat principle)}
{\it Let $z \in {\cal L}_{p,\gamma}^{+,\infty}$ such that (2.6) 
is satisfied. 
Then, $z$ is a critical point of 
$\tau$ if and only if it is a 
reparameterization of a $C^2$-geodesic in ${\cal L}_{p,\gamma}^{+,\infty}$.}

\1
\noindent
{\bf Proof.}
If $z$ is a critical point of $\tau$ (satisfying (2.6)) it is 
$$\int_{0}^{1} \langle D_{r}V,
{
{\dot z} \over {\langle U_z,\dot z \rangle}
}
\rangle dr = 0$$
for any $V$ along $z$ such that $V(0)=V(1)=0$. 
If $y$ is a reparameterization of $z$ such that 
$\langle U_w,\dot w \rangle$ is constant we deduce 
$\int_{0}^{1} \langle D_{r}V,\dot y \rangle = 0$ 
for any $V$ and therefore $y$ is a smooth geodesic.
Conversely is $z$ is a reparameterization of a $C^2$-geodesic,
it is $D_{s}(\mu' \dot z) = 0$ for some $\mu \in H^{1,\infty}([0,1],\R)$. 
Then $\langle U_{z},\mu' \dot z \rangle$ is constant, so 
$$
\int_{0}^{1} \langle D_{r}V,
{
{\dot z} \over {\langle U_z,\dot z \rangle}
}
\rangle dr = 0
$$ 
for any $V$ such that $V(0)=V(1)=0$.
Then, by Remark 2.8 we deduce that $z$ is a critical point of $\tau$.
\cvd

\1

\noindent
We are finally ready to prove Proposition 2.12.
\2
\noindent
{\bf Proof of Proposition 2.2.}
By Lemmas 2.5--2.6 and Theorem 2.9 there exists a future pointing
lightlike geodesic $w$ joining $q$ and $\gamma_\ast$, 
minimizing $\tau$ on
${\cal L}^{+,r}_{q,\ast}$. 
The uniqueness of $w$ is a consequence of the local invertibility of the
exponential map (cf. [19]).

\2

\noindent 
{\bf Remark 2.10.} 
Fix $z \in {\cal L}^{+,r}_{p,\gamma}(\Lambda)$ and 
$[a_1,a_2] \subset [0,1]$. 
Let $\gamma_2$ be the integral curve of $W$ such that
$\gamma_2(0)=z(a_2)$. 
Since it is future pointing, a simple contradiction argument shows that, 
whenever $d_R(z(a_1),z(a_2)) \longrightarrow 0$, the infimum of $\tau$ on
${\cal L}^{+,r}_{z(a_1),\gamma_2}([a_1,a_2],\Lambda)$ 
tends to $0$, and if $z(a_1)=z(a_2)$ the infimum is $0$.

\2
\noindent
{\bf Remark 2.11.} 
Since $\tau$ and ${\cal L}^{+,r}_{q,\gamma_*}(\Lambda)$ 
are invariant by reparameterizations, 
it is clear that there are nonsmooth minimizers. 
Note that, among the minimizers there are also curves having null 
derivatives in subsets of [0,1] with positive Lebesgue measure.
\1
Now we shall prove the equivalence between 
Definition 1.1 and the corresponding definition given in [9,10] 
(where the pseudocoercivity of $\tau$ is given in 
${\cal L}^{+,r}_{p,\gamma}(\Lambda)$).
\non
{\bf Lemma 2.12.} {\it Let $z\in {\cal L}^{+,1}_{p,\gamma}(\Lambda)$. 
Then there exists
$z_n$ in ${\cal B}^{+}_{p,\gamma}(\Lambda)$ such that $z_n\to z$ uniformly.}
\non
{\bf Proof.}  We apply can Proposition 2.2 to obtain the existence
of a minimizer in the space
${\cal L}^{+,1}_{z(a_1),\gamma_2}([a_1,a_2],\Lambda)$
where $\gamma_2$ is the integral curve of $w$ such that $\gamma_2(0)=z_2$.
Since $z$ is fixed, if $a_2-a_1$ is sufficiently small,
$\int_{a_1}^{a_2}{\sqrt{\langle \dot z,\dot z \rangle_{R}}}\,{\rm d}s$
is small and also the length (with respect to the Riemann structure (2.1))
of the geodesic minimizing $\tau$ is small. Then,
choosing a suitable partition of the interval $[0,1]$
allows to construct a broken geodesic $\hat z$ such that
the distance between $z$ and $\hat z$ with respect to the $H^{1,1}$-norm
is arbitrarily small.
Therefore, the uniform distance can be 
made as small as we want and we are done.
\cvd
\2

By Lemma 2.12 it follows immediately the following
\2
\noindent
{\bf Proposition 2.13.} 
{\it For any $r\in [1,+\io]$, $\tau$ is pseudo coercive
on ${\cal L}^{+,r}_{p,\gamma}(\Lambda)$ if and only if it is pseudo coercive
on ${\cal B}^+_{p,\gamma}(\Lambda)$.}
\1
For any $z\in H^{1,1}([0,1],\Lambda)$ denote 
by $\ell(z)$ its length induced by the Riemann structure (2.1).
\non
{\bf Lemma 2.14.} 
{\it Assume $\tau$ pseudocoercive on ${\cal B}^+_{p,\gamma}(\Lambda)$.
Then, for any $c\in\, ]\alpha,\beta[$ there exists $D(c)>0$ such that}
$$\tau(z)\le c\Rightarrow \ell(z)\le D(c)\ \ {\rm for\ any}\ \
z\in {\cal L}^{+,r}_{p,\gamma}(\Lambda).$$

\noindent
{\bf Proof.} 
Assume by contradiction the existence of a sequence $z_n$ in ${\cal
L}^{+,r}_{p,\gamma}(\Lambda)$ such that $\tau(z_n)\le c\in\,]\alpha,\beta[$
and $\ell(z_n)\to +\io$. Since $\tau$ and $\ell$ are invariant by
reparameterizations, by the pseudocoercivity of $\tau$ (and
Proposition 2.13) we can assume that $z_n$ has a subsequence (that we shall
continue to denote by $z_n$) uniformly convergent to a continuous curve
$z:[0,1]\to\bar\Lambda$. Since $z([0,1])$ is compact, it can be covered by a
finite number of open neighborhoods $U_i$ in $\cal M$ 
such that, on any $U_i$,
the metric $g$ has the form
$$ds^2=\langle\alpha(x,t)\xi,\xi\rangle_0
+2\langle\delta(x,t),\xi\rangle_0\theta-\theta^2$$
(cf. Remarks 2.3 and 2.4). 
If $]a_n^i,b_n^i[$ is an interval such that
$z_n(s)\in U_i$ for any $s\in ]a_n^i,b_n^i[$ and $z_n=(x_n,t_n)$ we have
$$
t_n(b_n^i)-t_n(a_n^i)=\int_{a^i_n}^{b_n^i} \dot t_n{\rm d}s
=\int_{a^i_n}^{b_n^i}\left(
\langle \delta,\dot x_n\rangle_0+\sqrt{\langle \delta,\dot x_n\rangle_0^2+
\langle \alpha\dot x_n,\dot x_n\rangle_0}
\right){\rm d}s .
$$
Since any $\bar U_i$ is compact and the number of $U_i$ is finite we deduce
that $\int_{a^i_n}^{b_n^i} \sqrt{\langle\dot x_n,\dot x_n\rangle_{0}}ds$ (and
consequently $\int_{a^i_n}^{b_n^i}|\dot t_n|ds)$ is bounded independently by
$n$ and $i$. 
Finally, since any $z_n$ is included in $\bigcup\limits_i U_i$ and $\dot
z_n=(\dot x_n,\dot t_n)$ on any $U_i$, we deduce that
$\int_{a^i_n}^{b_n^i} \|\dot z_n\|_{R}ds$ is bounded independently by $n$ and
$i$. Then $\ell(z_n)$ is bounded getting a contradiction.\cvd
\2
\noindent
Lemma 2.14 will be used, together with the following proposition, to
construct the shortening flow for $\tau$.
\non
{\bf Proposition 2.15.} 
{\it Let $W$, $\Lambda$ and $\gamma:]\alpha,\beta[\to
\Lambda$ be satisfying {\rm (1)--(6)} of {\rm Sect. 1}. 
Assume that $\tau$ is
pseudocoercive on ${\cal B}^+_{p,\gamma}(\Lambda)$ and fix $c\in
]\alpha,\beta[$.
Then there exists $\rho_*(c)>0$ satisfying the following property. Let
$z\in\tau^c\cap{\cal L}^{+,r}_{p,\gamma}(\Lambda)$, 
$[a,b]\subset [0,1]$, $z_1,z_2\in z([0,1])$ with $z_1=z(a),\
z_2=z(b)$ and $d_R(z_1,z_2)\le \rho_*(c)$. Let $\gamma_i : ]\alpha_{i}^{-},
\beta_{i}^{+}[ \longrightarrow \Lambda (i=1,2)$ be the maximal
integral curve (in $\Lambda$) of
$$\cases{ \dot\eta=W(\eta)\cr
\eta(0)=z_i, \quad i=1,2.} \eqno(2.8)$$
Moreover, 
for any $\hat z_1 \in \gamma(]\alpha_{1}^{-},0])$ 
with $d_R(\hat z_1,z_1) \leq \rho_{*}(c)$
there exists a unique future pointing light--like geodesic
$\Gamma$ such that $\Gamma(a)=\hat z_1,\ \Gamma(b)$ is in the image of
$\gamma_2$, $\Gamma(s)\in\Lambda$ for any $s\in [a,b]$ and
$$\tau(\Gamma)=\inf\Big\{\tau(y):y\in{\cal
L}^{+,r}_{\hat z_1,\gamma_2}([a,b],\Lambda)\Big\}\eqno(2.9)$$}
\non
{\bf Proof.} By pseudo coercivity it is immediate to
check the existence of $K$, compact subset of $\bar\Lambda$, such that
$$z\in \tau^c \Rightarrow z([0,1])\subset K\eqno(2.10)$$
Now take a finite family $U_1,\ldots , U_m$ of open subsets of $\cal M$
covering $K$ and such that any $U_i$ is compact and $U_i$ satisfies the
properties of Remarks 2.3 and 2.4. 
By Proposition 2.2 and Remark 2.10 if $\rho_*(c)$ is
sufficiently small there exists a minimizer $w$ in ${\cal
L}^{+,r}_{\hat z_1,\gamma_2}([a,b],\ U_i)$ for some $i=1,\ldots,m$. 
(Note that, by assumption (6),
$w(b) \in \Lambda).$ 
Since $\bigcup\limits_{i=1}^{m}$ $\bar U_i$ is compact, 
by the local invertibility of the exponential map 
(and the minimality of $\tau(w)$) 
we see that, if $\rho_*(c)$ is sufficiently small, 
the minimizing geodesic is unique. Then we have 
just to prove that the minimizer is included in $\Lambda$. If $z_1=z_2$
this is obvious. Then suppose that $z_1\ne z_2$. If  $\rho_*(c)$
is sufficiently small and
$d_R(z_1,z_2) \leq \rho_{*}(c)$,  
$z_1,z_2\in U_{i}\cap \Lambda$ (for some i). 
Then, using Proposition 2.2 and choosing $\rho_*(c)$ sufficiently small,
we can construct two continuous map
$\theta_1,\theta_2:[0,1]\to U_i\cap\Lambda$ having the
following properties:
\vs
\item{$\bullet$} for any $\lambda\in [0,1],\ \theta_2(\lambda)$ is in the
future of $\theta_1(\lambda)$
\vs
\item{$\bullet$} $\theta_1(0)=\hat z_1\ \theta_2(0)=z_2$
\vs
\item{$\bullet$} $\theta_1(\lambda)\ne\theta_2(\lambda)$ for any
$\lambda\ne 1$.
\vs
\item{$\bullet$} $\theta_1(1)=\theta_2(1)$.
\vs
\item{$\bullet$} for any $\lambda\in [0,1]$ there exists a unique minimizer
of $\tau$ on
${\cal L}_{\theta_1(\lambda),\gamma_2(\lambda)}([a,b],U_i)$ where
$\gamma_2(\lambda)$ is the maximal integral curve of $W$ such that
$\gamma_2(\lambda)(0)=\theta_2(\lambda)$.
\vs 

Now set
$$\eqalign{ 
A=\Big\{\lambda\in & [0,1]:\ \hbox{\it the light--like  or constant geodesic
minimizing $\tau$ on} \cr
& {\cal L}_{\theta_1(\lambda),\gamma_2(\lambda)}([a,b],U_i)\ \ 
\hbox{\it does not intersect}\ \ \partial\Lambda \Big\}.}
$$
Since $\theta_1(1)=\theta_2(1)\in\Lambda,\ 1\in A$. Take 
$$
\lambda_0\equiv\inf A\ge 0.
$$
By the definition of $\lambda_0$, there exists $\lambda_n\to\lambda_0^+$ and
a sequence $w_n$ of lightlike geodesic minimizing $\tau$ on
${\cal L}^{+,r}_{\theta_1(\lambda_n),\gamma_2(\lambda_n)}([a,b],U_i)$ such
that $w_n([0,1])\subset \Lambda$. 
Unless to consider a subsequence, 
by (2.4) we obtain  the existence of a lightlike geodesic $w$ such that
$$\eqalign
{& w_n\to w\ \ {\rm with\ respect\ to\ the}\ \ C^2-{\rm norm}\cr
& w(a)=\theta_1(\lambda_0),\ \ w(b)
\in \gamma_2^{-1}(\lambda_0) \quad(\gamma_2 \subset \Lambda)\cr
& w([a,b])\subset\bar\Lambda.}$$
If $w([a,b] \subset \Lambda$, then $\lambda_0 = 0$ and we are done. 
If $w([a,b]) \cap \partial\Lambda \not= \emptyset$,
since $\theta_1(\lambda_0)\ne\theta_2(\lambda_0)$ are in $\Lambda$ we get a
contradiction with the light--convexity of $\bar\Lambda$.\cvd
\2
\centerline{\bf 3. HOMOTOPICAL EQUIVALENCE BETWEEN ${\cal
L}^{+,r}_{p,\gamma}(\Lambda)$ AND ${\cal B}^+_{p,\gamma}(\Lambda)$}
\1
In this section (under assumptions (1)--(6)) we shall introduce a 
shortening flow and we shall
use it to prove that ${\cal L}^{+,r}_{p,\gamma}(\Lambda)$
and ${\cal B}^+_{p,\gamma}(\Lambda)$ are homotopically equivalent for 
any $r\in [1,+\infty]$. 
This flow will also be used to get the deformation of the
sublevels of $\tau$ for the Morse Theory, whenever we are far from
lightlike geodesics. 
Indeed far from lightlike geodesics, 
$\tau$ will be strictly decreasing with a speed uniformly far from zero. 
In other words $\tau$ will verify the Palais-Smale compactness condition 
along the flow.
To construct the shortening flow we shall use the same ideas in
[18] adapting them to our case. 
Note that here we can not use the finite dimensional approach 
nearby critical curves (used in [18] for Riemannian geodesics) 
because we are not working with fixed endpoints. 
So the shortening approach will be used only far from geodesics.
The shortening procedure can be introduced in the following way.
Fix $c > \inf\{\tau(z), z \in {\cal L}^{+,r}_{p,\gamma}(\Lambda) \}$.
Consider $D(c)$ as in Lemma 2.14, $\rho_*(c)$ as in
Proposition 2.15 and take $N = N(c)$ such that
$$
{
{D(c)}
\over
N
}
< \rho_*(c)\die .
$$
Choose a partition $\{0 = s_0 < s_1 \dots s_{N-1} < s_N =1\}$ of $[0,1]$
such that for any $i \in \{1, \dots N\}$,
$$
s_{i} - s_{i-1} = {1\over N}\die .
$$
For any $z \in \tau^c\cap {\cal L}^{+,r}_{p,\gamma}(\Lambda)$, 
choose $N+1$ points
$z_0,z_1, \dots z_{N}$ on $z([0,1])$ such that $z(0) = p$, $z_N = z(1)$ and
$d_R(z_i,z_{i-1}) = l(z)/N$, for any $i \in \{1, \dots N\}$, where $l(z)$
denotes the length of $z$ with respect to the Riemannian structure (2.1)
(see Figure~1).
Denote by $\gamma_i$ ($i = 1,\dots,N)$ the maximal integral
curve of $W$ such that $\gamma_i(0) = z_i$ (see Figure~2).
Observe that $\gamma_N(s)=\gamma(s+\tau(z))$ for all $s$.
Let $w_1$ be the lightlike geodesic minimizing $\tau$ on
${\cal L}^{+,r}_{p,\gamma_1}([s_0,s_1],\Lambda)$ 
(recall that $z_0 = p$ and $s_0=0$),
$w_2$ the lightlike geodesic minimizing $\tau$ on
${\cal L}^{+,r}_{w_1(s_1),\gamma_2}([s_1,s_2],\Lambda)$, and so on 
(see Figure~3).
In Figures 3, 4 and 5 the points $w_i(s_i)$ are denoted by
$\overline w_i$.
Note that the number $N$ can be chosen big enough in order that
$d_R(w_i(s_i),z_{i+1}) \leq \rho_*(c)$, for any $i=1,\dots,N-1$ 
and for any $z\in\tau^c$.
\2
\noindent
{\bf Remark 3.1.}~
Let $K=K(c)$ be a compact subset of $\bar \Lambda$ as in (2.10). 
By compactness, $K(c)$ can be covered by a finite family
$(U_j)$ satisfying Remark 2.3. 
Moreover, $N$ can be chosen so large that
$z([s_{i-1},s_i])$ and the minimizer of $\tau$ on
${\cal L}^{+,r}_{w_{i-1}(s_{i-1}),\gamma_i}([s_{i-1},s_i],\Lambda)$
are contained in some $U_j$.
\2
The Lorentzian metric on $U_j$ is described as
$$
\langle \zeta,\zeta\rangle =
\langle\alpha_j(x,t)\xi,\xi\rangle_0 +
2\langle\delta_j(x,t),\xi\rangle_0\theta -\theta^2~
\eqno(3.1)
$$
(cf. Remark 2.3), 
where $\alpha_j(x,t)$ is a positive linear operator, $\delta_j(x,t)$ is
smooth vector field, $z = (x,t) \in U_j$ and $\zeta = (\xi,\theta) \in T_z\M$.
With the notation above, 
for any  future pointing curve $z$ with image contained in some
$U_j$,  the condition
$\langle\dot z,\dot z\rangle = 0$ holds if and only if
$$
\dot t =
{\langle\delta_j(x,t),\dot x\rangle}_0
+
\sqrt{
{\langle\alpha_j(x,t),\dot x,\dot x\rangle}_0
+
{\langle\delta_j(x,t),\dot x\rangle}^2_0
}
\eqno(3.2)
$$
Moreover, any $\gamma_i$ is an integral curve of $W$, so, in $U_j$, it
has the form $s \longmapsto (x_j,t{_j}+s)$, if $z_j=(x_j,t_j)$.
Note that
${\cal L}^{+,r}_{p,\gamma_1}([s_0,s_1],\Lambda)$
is nonempty, since it contains the restriction $z_{|[s_0,s_1]}$.
Now, using elementary comparison theorems for ordinary differential equations
and the metric (3.1) on $U_j$ allow to deduce that also any
space
${\cal L}^{+,r}_{w_{i-1}(s_{i-1}),\gamma_{i},\ep}([s_{i-1},s_i],\Lambda)$
is nonempty for
any $i \in \{2,\dots N\}$.

Note also that, if $\eta_1$ is the curve defined by setting
$\eta_1 ([s_{i-1},s_i]) = w_i$, 
then $\tau(\eta_1) \leq \tau(z)\le c$ (always by
comparison theorems in O.D.E.).
In particular $\eta_1([0,1])$ is contained in $K(c)$.

\2
\noindent
{\bf Remark 3.2}~
A second curve $\eta_2$ will be constructed
in the following way starting from $\eta_1$.
On any minimizer $w_i$ ($i=1,\dots,N$) consider the point $m_{i}$
such that
$d(w_i(s_{i-1}), m_i) = d(m_i,w(s_i))$.

For $i =1,\dots,N$, 
we denote by $\lambda_i$ the maximal integral curve of $W$ such that
$\lambda_i(0) = m_i$; moreover, we set $\lambda_{N+1}(s) =
\gamma(s+\tau(\eta_1))$ (see Figure~4).

Consider now the following subdivision of the interval $[0,1]$. Let
$\sigma_0=0$, $\sigma_1={1\over 2N}$, 
$\sigma_j={2j-1\over 2N}$ for $j=2,\ldots,N$,
and $\sigma_{N+1}=1$.

Denote by $u_1$ the minimizer of $\tau$ on 
${\cal L}^{+,r}_{p,\lambda_1}([\sigma_0,\sigma_1],\Lambda)$,
by $u_2$ the minimizer of $\tau$ on
${\cal L}^{+,r}_{u_1(\sigma_1),\lambda_2}([\sigma_1,\sigma_2],\Lambda)$ 
and so, inductively,
we denote by $u_j$ the minimizer of $\tau$ in 
${\cal L}_{u_{j-1}(\sigma_{j-1}),
\lambda_{j},\ep}([\sigma_{j-1},\sigma_j],\Lambda)$, $j=2,\ldots,N+1$.
Finally, (see Figure~5) we denote by
$\eta_2$ the curve such that
$$
\eta_2\vert_{[\sigma_{j-1},\sigma_j]}=u_j~.
$$
Using again comparison theorems in ordinary differential equations
one proves that
$\tau(\eta_2) \leq \tau(\eta_1)$.
\2
The continuous flow $\eta(\sigma,z)$ can be constructed as follows.
Fix $\sigma \in [0,1]$ and consider for instance the interval
$[s_0,s_1]$.
We choose $\eta(\sigma,z)_{|[s_0,s_1]}$ as follows.
Set $p = (x_0,0)$ and $\gamma_1(s) = (x_1,t_1+s)$
(in some neighborhood $U_j$ as in Remark 3.1).
Since $z(s) = (x(s),t(s))$, the curve $x(s)$ joins $x_0$ with $x_1$.

Let $y(\sigma)$ be the minimizer of the functional
$$\displaylines{
y \longmapsto
\int_{s_0}^{\sigma s_1}
\langle\delta_i(y,t_y),\dot y\rangle_0 
{\rm d}s
+
\cr
\int_{s_0}^{\sigma s_1}
\sqrt{
\langle\alpha_i(y,t_y),\dot y,\dot y\rangle_0
+
\langle\delta_i(y,t_y),\dot y\rangle_0^2 
}
\cr}
$$
with boundary conditions $y(0) = x_0$ and $y(\sigma s_1) = x(\sigma
s_1)$,
where $t_y$ is the solution of (3.2) with 
$t_y(0) = 0$ in the interval $[0,{\sigma s_1}]$.
Denote by $\hat y(\sigma)$ the extension of $y(\sigma)$ to $[s_0,s_1]$
taking $\hat y(s) = x(s)$ for $s \in [\sigma s_1,s_1]$.
Finally, denote by $\hat t_y$ the corresponding solution of (3.2) in the
interval $[s_0,s_1]$.
The curve $(\hat y(\sigma),\hat t_y(\sigma))$ will be $\eta(\sigma,z)$ in the
interval $[s_0,s_1]$.
In the same way we can construct $\eta(\sigma,z)$ on the other intervals
$[s_{i-1},s_i]$. Note that, by construction,  $\eta(1,z)=\eta_1$.
Similarly, we can extend the flow $\eta$ to a map defined
on $[0,2]\times\tau^c$ in such a way that $\eta(2,z)=\eta_2$.
Now, we iterate the shortening argument above,  
replacing the original curve $z$ with
the curve $\eta_2$.
Successively we apply the above construction, starting from $\eta_2$.
By induction we
obtain a flow $\eta(\sigma,z)$, defined on $\R^+\times\tau^c$.
Since $\tau(\eta(\sigma,z))\le\tau(z)$ for any $\sigma$ and for any $z$, 
using $\eta$ we immediately deduce 
\non
{\bf Lemma 3.3.} {\it Fix $r\in [1,+\infty].$ 
For any $c\in \, ]\alpha,\beta[\,$,
${\cal L}^{+,r}_{p,\gamma}(\Lambda)\cap \tau^c$ is homotopically
equivalent to ${\cal B}^+_{p,\gamma}\cap \tau^c$.}
\2

Moreover choosing a suitable continuous map 
$\rho_*(c)$ ($c \in ]\alpha,\beta[$) 
and arguing as in section 9 of [8] we can also obtain
\non
{\bf Proposition 3.4.} 
{\it ${\cal L}^{+,r}_{p,\gamma}(\Lambda)$ is homotopically
equivalent to ${\cal B}^+_{p,\gamma} (\Lambda)$.}
\1
Suppose that $\tau(\eta_1)=\tau(\eta_2)$ 
and consider the situation is a single interval $[\sigma_j,\sigma_{j+1}]$.
Since $\tau(\eta_1)=\tau(\eta_2)$ 
simple comparison theorems in O.D.E. show that 
$\eta_1$ is a minimizer on the interval $[\sigma_j,\sigma_{j+1}]$.
Suppose that it consists of two (nonconstant) lightlike geodesics. 
If it is not a lightlike geodesic, 
by the above construction it has a discontinuity at 
$s_{j+1} = 
{
{\sigma_{j+1} + \sigma_{j}}
\over
{2}
}
$.
Denote by $U_{\eta_{1}}$ the parallel transport of 
$\dot \gamma(\tau(\eta_{1}))$ 
along the curve $\eta_1$. 
Since $\eta_1$ is a minimizer satisfying (2.6), by Lemma 2.7
and Remark 2.8 it is 
$$
\int_{\sigma_j}^{\sigma_{j+1}}
{\langle D_{s}V,{\dot \eta_1 \rangle} 
\over 
{\langle U_{\eta_{1}},\dot \eta_1 \rangle}} ds = 0
$$
for any $C^{\io}$-vector field along $\eta_1$ such that $V(0)=0, V(1)=0$.
In particular 
${\dot \eta_1} \over {\langle U_{\eta_{1}},\dot \eta_1 \rangle}$ is a
$C^1$ curve. Therefore $
\dot \eta_{1}(s_{j+1}^{-})=\dot
\eta_{1}(s_{j+1}^{+})$, so $\eta_1$ is the image of a future 
pointing lightlike geodesic in the interval $[\sigma_j,\sigma_{j+1}]$. 

Then, whenever we are far from lightlike geodesics and there are 
not intervals where $\eta_1$ is  a constant , $\tau(\eta_2) < \tau(\eta_1)$. 
If  $\eta_1$ possesses some interval where it is a constant it 
is possible to construct a 
"localized" flow where $\tau$ is strictly decreasing 
ignoring such intervals and using the above 
construction of the flow in a small neighborhood of $\eta_1$. 

Finally compactness arguments similar to the ones used for 
the shortening method for Riemannian geodesics (cf. [18]), allows to obtain  
the analogous of the classical deformation results (cf e.g. [4,17])  
for the functional $\tau$ on 
${\cal L}^{+,r}_{p,\gamma}(\Lambda)$. 

Since, to obtain Morse Relations, we shall work with respect 
to the $H^{1,2}$ structure , 
we give the statements of the deformation results only
for $r=2$.

\2
\noindent
{\bf Proposition 3.5.}~
{\it
Let $c$ be a regular value for $\tau$ on
${\cal L}^{+,2}_{p,\gamma}$ (namely $\tau^{-1}(\{c\})$ does not 
contain geodesics). 

Then, there exists a positive number $\delta = \delta (c)$ and a continuous
map
$H \in C^0([0,1]\times \tau^{c+\delta},\tau^{c+\delta})$, such that:}

\smallskip

\item{(a)} $H(0,z) = z$, {\it for every $z \in  \tau^{c+\delta}$;}
\smallskip
\item{(b)} $H(1,\tau^{c+\delta}) \subseteq \tau^{c-\delta}$;\smallskip
\item{(c)} $H(\sigma,z) \in \tau^{c-\delta}$,
{\it for any $\sigma\in [0,1]$ and $z \in \tau^{c-\delta}$};
\smallskip
\2
\noindent
{\bf Proposition 3.6}~
{\it 
Let $K_c$ be the set of  lightlike geodesics on 
${\tau}^{-1}(\{c\}) \cap {\cal L}^{+,2}_{p,\gamma}$.
Then for any open neighborhood $U$ of $K_c$, there exists a positive number
$\delta = \delta(U,c)$ and a homotopy
$H \in C^0([0,1]\times \tau^{c+\delta},\tau^{c+\delta})$, such that}
\smallskip
\item{(a)} $H(0,z) = z$, {\it for any $z \in  \tau^{c+\delta}$.}
\smallskip
\item{(b)}
$H(1,\tau^{c+\delta} \setminus U) \subset \tau^{c-\delta}$;
\smallskip
\item{(c)} $H(\sigma,z) \in \tau^{c-\delta}$,
{\it for every $\sigma\in [0,1]$ and $z \in \tau^{c-\delta}$}.
\medskip\noindent
\2
\noindent
{\bf Remark 3.7.} 
\enspace There are two main differences
between the shortening method described above
and the classical shortening method for Riemannian
geodesics.  In our case, we locally minimize a functional which is
is not given in an integral form.
Secondly, we minimize the functional in the space of curves joining
a point with a curve, and not two fixed points.

\non
{\bf Remark 3.8.} 
The flow used in proving 
Propositions 3.5 and 3.6 are just what 
we need for a Ljusternik--Schnirelmann theory. 
Then, without using the nondegeneracy
assumption of Theorem 1.5 we can obtained the existence of at last 
$cat ({\cal B}^+_{p,\gamma}(\Lambda))$ future pointing light--like geodesic in
${\cal B}^+_{p,\gamma}(\Lambda)$. (Here cat $X$ denotes the minimal number of
contractible subsets of $X$ covering it). Moreover if 
$cat ({\cal B}^+_{p,\gamma}(\Lambda))=+\infty$ 
there is a sequence $z_n$ of future
pointing light--like geodesics in ${\cal B}^+_{p,\gamma}(\Lambda)$
such that
$\tau (z_n)\to \beta$.

\2

\centerline{\bf 4. ON THE BEHAVIOUR OF $\tau$ NEARBY LIGHTLIKE GEODESICS}\vs
\1

To develop a Morse Theory we shall use the space $\L$ (that contains 
${\cal B}^{+,2}_{p,\gamma}(\Lambda)$), because $H^{1,2}([0,1],\Lambda)$
is an Hilbert manifold (endowed with its natural metric). 
Since $\L$ is invariant by reparameterization
as well as $\tau$, it will be useful to consider equivalence classes
of curves or, better, to single out one parameterization.
This will be done on a open neighborhood ${\cal N}_{w}$ of $w([0,1])$
for any lightlike geodesic $w$.

Towards this goal consider the parallel vector field 
$U_w$ along $w$ of $\dot \gamma(\tau(w))$ (which is a time-like vector).
Since $\gamma$ is an integral curve of $W$,
by Remark 3.4 it is $\langle U_w,U_w \rangle \equiv -1$. 
Now, by the pseudocoercivity of $\tau$ it follows that 
$w$ does not have self-intersection, so its image is a submanifold and 
$U_w$ can be extended to a smooth vector field $Y$ on $\Lambda$ such that 
$$
\langle Y(z),Y(z) \rangle = -1 \hbox{  for any  } z \in \Lambda,
\eqno(4.1)
$$
$$
D_{\dot w(s)}Y(w(s)) = 0 \hbox { for any  } s \in [0,1],
\eqno(4.2)
$$
(cf. [30]). Using the vector field $Y$ we define the following space:
$$
\Q = \Big\{z \in \L :
\langle Y(z),\dot z \rangle = 
\int_{0}^{1}{\langle Y(z),\dot z \rangle}ds \hbox { a.e. } \}.
\eqno(4.3)
$$
Note that by (4.2), the geodesic $w$ is in $\Q$ and 
$$
 \int_{0}^{1}{\langle Y(w),\dot w \rangle}ds < 0.
$$
The space $\Q$ will be used to study 
Morse Theory nearby the critical point $w$ for the functional $\tau$. 
The first step in this direction is to prove that $\Q$ is a $C^1$-manifold.

\2
\noindent
{\bf Remark 4.1.}  
It is well known that the space 
$\Om$ defined by (2.2) is a 
$C^{\io}$-manifold and its tangent space at any $z$ is given by 
$$\displaylines{
T_{z}\Om = \Big\{\zeta \in H^{1,2}([0,1],{\cal T}{\Lambda}): \cr
\zeta (s) \in T_{z(s)}\Lambda \hbox{  for any $s$,  }
\zeta(0)=0, \die \zeta(1))
\hbox { is parallel to  } \dot\gamma(\tau(z)) \} \cr}
$$
where ${\cal T}{\Lambda}$ denotes the tangent bundle of $\Lambda$.
\2
\noindent
{\bf Remark 4.2.} Consider the map 
$$\phi : \Om \longrightarrow \Big\{h \in L^2([0,1],\r) : 
\int_{0}^{1}h\die ds=0\}$$ 
defined as
$$
\phi(z) = \langle Y(z),\dot z \rangle - 
\int_{0}^{1}{\langle Y(z),\dot z \rangle}ds.
\eqno(4.4)
$$
It is a standard computation to prove that 
$\phi$ is of class $C^\io$ and its differential satisfies:
$$\displaylines{
d\phi(z)[\zeta] = 
\langle Y,D_{s}\zeta \rangle + \langle D_{\zeta}Y,\dot z \rangle + \cr 
\hfill 
-\int_{0}^{1}({\langle Y,D_{s}\zeta \rangle + 
\langle D_{\zeta}Y,\dot z \rangle})ds \hfill \llap(4.5) \cr}
$$
where $D$ is the Levi-Civita connection relatively to the Lorentzian 
structure $g$.

\2
\noindent
{\bf Proposition 4.3.} {\it The space 
$$
\P = \Big\{z \in \Om : 
\langle Y(z),\dot z \rangle = \
\int_{0}^{1}{\langle Y(z),\dot z \rangle}ds < 0 \},
\eqno(4.6)
$$
is a manifold whose tangent space is given by}
$$\displaylines{
T_{z}\P = \Big\{\zeta \in T_{z}\Om : 
\langle Y,D_{s}\zeta \rangle + \langle D_{\zeta}Y,\dot z \rangle = \cr
\hfill \int_{0}^{1}({\langle Y,D_{s}\zeta \rangle +
\langle D_{\zeta}Y,\dot z \rangle})ds \hfill \llap(4.7) \cr}
$$

\1
\noindent
{\bf Proof.} 
Consider the map $\phi$ defined by (4.4). 
By (4.5) and the Implicit Function Theorem it 
is sufficient to prove that  for any 
$h \in L^2([0,1],\r)$ such that 
$\int_{0}^{1}h\die ds=0$ there exists $\zeta \in T_{z}\Om$ such that 
$$ 
\langle Y,D_{s}\zeta \rangle + \langle D_{\zeta}Y,\dot z \rangle = h.
\eqno(4.8)
$$
Choose $\zeta = \mu Y$ with $\mu(0)=0$. Then $\zeta$ satisfies
(4.8) if and only if $\mu$ satisfies the Cauchy problem
$$
\cases {-\dot \mu + \mu \langle D_{Y}Y,\dot z \rangle = h, \cr
\mu(0)=0,\cr}
$$
because, by (4.1), 
$\langle Y,Y \rangle \equiv -1$. 
Such a problem has a (unique) solution in $H^{1,2}([0,1],\r)$ and we are done.
\cvd

\2
\noindent
{\bf Remark 4.4.} 
For any $z \in \P$, ${\Vert \dot z \Vert}_R$ is uniformly far from $0$. 
(Here  ${\Vert \cdot \Vert}_R$ is the norm induced by the 
Riemann structure (2.1)). Indeed if $z \in \P$ it is 
$$\displaylines{
0 < -\int_{0}^{1}{\langle Y(z),\dot z \rangle}ds = 
-{\langle Y(z),\dot z \rangle} \leq \cr
\mid {\langle Y(z),\dot z \rangle}_{R} 
\mid \leq {\Vert Y(z) \Vert}_{R} {\Vert \dot z \Vert}_{R}. \cr}
$$

\2
\noindent
{\bf Lemma 4.5.} {\it Let $\psi : \P \longrightarrow L^{2}([0,1],\r)$
defined as}
$$
\psi(z)=\sqrt{2}{\langle \dot z, W(z) \rangle} + 
\sqrt{\langle \dot z,\dot z \rangle_{R}}
\eqno(4.9)
$$
{\it Then $\psi$ is of class $C^1$ and, for any $\zeta \in \P$,}
$$\displaylines{
d\psi(z)[\zeta] = 
\sqrt{2} (\langle D_{s}\zeta,W(z) \rangle + 
\langle D_{\zeta}W,\dot z \rangle)+ \cr
\hfill
{{1} \over {\sqrt{{\langle \dot z,\dot z \rangle}_{R}}}}
\left( \langle D_{s} \zeta,\dot z \rangle +2 \langle \dot z,W(z) \rangle
( \langle D_{s} \zeta,W(z) \rangle + 
\langle D_{\zeta}W,\dot z \rangle) \right).
\hfill \llap(4.10) \cr}
$$
Note that, by Remark 4.4, 
since $\zeta \in H^{1,2}$ it is $d\psi(z)[\zeta] \in L^{2}$.

\1
\noindent
{\bf Proof. } Standard computations show that the differential of the map 
$$
\psi_{1}(z) = \sqrt{2} {\langle \dot z, W(z) \rangle}
$$
along the direction $\zeta$ is given by
$$
d\psi_{1}(z)[\zeta] = 
\sqrt{2} 
(\langle D_{s}\zeta,W(z) \rangle + \langle D_{\zeta}W,\dot z \rangle).
$$
To evaluate the differential of the map
$$
\psi_{2}(z)=\sqrt{\langle \dot z,\dot z \rangle_{R}}$$
(at an instant $s_0$) 
we can assume that (in a neighborhood of $z(s_{0})$ we are in $\r^n$ and 
$$
{\langle \zeta,\zeta \rangle}_{R} = {\langle L(z)[\zeta], \zeta \rangle}_{E}
$$
where ${\langle \cdot, \cdot \rangle}_{E}$ is the Euclidean scalar product
of $\r^n$ and $L(z)$ is a smooth positive definite linear operator. Using
such a position the vector fields in the tangent space at any curve $z$
(on a suitable interval $[s_{0}-\delta,s_{0}+\delta]$) will be
$H^{1,2}$-vector fields 
(defined on $[s_{0}-\delta,s_{0}+\delta])$  with values in $\r^n$.

Suppose that $\zeta$ is of class $C^1$. Then, by Remark 4.4,
$\dot z + \lambda \dot{\zeta}$ is uniformly far from $0$ for any
$\lambda$ sufficiently small, and
$$
d{\psi}_{2}(z)[\zeta] =
\displaystyle\lim_{\lambda \to 0}
{{1 \over \lambda}
( \sqrt{\langle L(z+{\lambda}{\zeta})[\dot z + {\lambda}{\dot \zeta}],
\dot z + {\lambda}{\dot \zeta} \rangle_{E}}
- \sqrt{\langle L(z)[\dot z],\dot z \rangle _{E} })}. 
$$
Therefore there exists $\theta = \theta(\lambda,s) \in [0,1]$ such that 
$$
d{\psi}_{2}(z)[\zeta] =
\displaystyle\lim_{\lambda \to 0}
{
{{\langle dL(z+{\lambda}{\theta}{\zeta})[\zeta]
[\dot z +{\lambda}{\dot \zeta}], \dot z +{\lambda}{\dot \zeta} \rangle_{E}
+ 2 \langle L(z +{\lambda}{\theta}{\zeta})[\dot \zeta],  \dot z + {\lambda}
{\theta}{\dot \zeta} \rangle_{E}}
\over
{2\sqrt{\langle L(z+{\lambda}{\theta}{\zeta})
[\dot z +{\lambda}{\theta}{\dot \zeta}],\dot z + {\lambda}{\theta}{\dot \zeta}
\rangle_{E} }}}
} 
$$
where the limit is done with respect to the $L^2$ norm.

Then, by the Lebesgue Convergence Theorem we obtain
$$
d{\psi}_{2}(z)[\zeta] =
{{\langle dL(z)[\zeta]
[\dot z], \dot z \rangle_{E}
+ 2 \langle L(z)[\dot \zeta],  \dot z \rangle_{E}}
\over
{2\sqrt{\langle L(z)[\dot z],\dot z \rangle_{E} }}}
$$
which is a map in $L^{2}([s_{0}-\delta,s_{0}+\delta])$ by Remark 4.4.

Since 
$$
\langle L(z)[\dot z],\dot z] \rangle_{E} = 
\langle \dot z, \dot z \rangle_{R} =
\langle \dot z, \dot z \rangle + 2{\langle W(z),\dot z \rangle}^{2}
$$
and 
$$
\langle dL(z)[\zeta]
[\dot z], \dot z \rangle_{E}
+ 2 \langle L(z)[\dot \zeta],  \dot z \rangle_{E}
= d(\langle L(z)[\dot z],\dot z \rangle_{E})[\zeta],
$$
we obtain
$$
d{\psi}_{2}(z)[\zeta] =
{{1} \over {\sqrt{{\langle \dot z,\dot z \rangle}_{R}}}}
\left( \langle D_{s} \zeta,\dot z \rangle +2 \langle \dot z,W(z) \rangle
( \langle D_{s} \zeta,W(z) \rangle + 
\langle D_{\zeta}W,\dot z \rangle) \right).
\eqno(4.11)
$$

Now consider $\zeta \in H^{1,2}([s_{0}-\delta,s_{0}+\delta],\r^n)$ 
and $\zeta_{1}$ of class $C^1$. 
Then (in local coordinates)
for any $\lambda$ sufficiently small with respect to 
$\Vert \dot \zeta_{1} \Vert_{L^\io}$, there exist $C_1,C_2 > 0$ such that 
$$\displaylines{
\left| 
{
{\psi_2(z+\lambda\zeta_{1})-\psi_2(z+\lambda\zeta)}
\over
{\lambda}
}
\right| = \cr
{1 \over \lambda}
\left|
{
{\langle L(z+{\lambda}{\zeta_1})[\dot z + {\lambda}{\dot \zeta_1}],
\dot z +{\lambda}{\dot \zeta_1} \rangle_{E}
-\langle L(z+{\lambda}{\zeta})[\dot z + {\lambda}{\dot \zeta}],
\dot z +{\lambda}{\dot \zeta} \rangle_{E}}
\over
{\sqrt{\langle L(z+{\lambda}{\zeta_1})
[\dot z +{\lambda}{\dot \zeta_1}],\dot z + {\lambda}{\dot \zeta_1}
\rangle_{E}} 
+
\sqrt{\langle L(z+{\lambda}{\zeta})
[\dot z +{\lambda}{\dot \zeta}],\dot z + {\lambda}{\dot \zeta}
\rangle_{E}}}
}
\right| \cr
\leq
{
{C_{1} \Vert \zeta -\zeta_{1} \Vert_{L^\io} 
\langle \dot z +{\lambda}{\dot \zeta_1},\dot z + 
{\lambda}{\dot \zeta_1} \rangle_{E}}
\over
{\sqrt{\langle L(z+{\lambda}{\zeta_1})
[\dot z +{\lambda}{\dot \zeta_1}],\dot z + {\lambda}{\dot \zeta_1}
\rangle_{E}}}
}+ \cr
{
{C_{2}| \dot z +{\lambda}{\dot \zeta_1} + \dot z + {\lambda}{\dot \zeta}|_{E}
|\zeta_{1}-\zeta|_{E}}
\over 
{\sqrt{\langle L(z+{\lambda}{\zeta_1})
[\dot z +{\lambda}{\dot \zeta_1}],\dot z + {\lambda}{\dot \zeta_1}
\rangle_{E}} 
+
\sqrt{\langle L(z+{\lambda}{\zeta})
[\dot z +{\lambda}{\dot \zeta}],\dot z + {\lambda}{\dot \zeta}
\rangle_{E}}}
}
\cr }
$$
where $| \cdot |_{E}$ is the 
Euclidean norm in $\r^n$. Then there exists $C > 0$ such that 
$$
\left| 
{
{\psi_2(z+\lambda\zeta_{1})-\psi_2(z+\lambda\zeta)}
\over
{\lambda}
}
\right| 
\leq
C(\Vert \zeta_1 - \zeta \Vert_{L^\io} |\dot z + 
{\lambda}{\dot \zeta_1}|_{E} + |\dot \zeta_{1} - \dot \zeta|_{E}).
$$
Therefore, since $\dot z$ is in $L^2$, 
$\Vert \dot \zeta_{1} \Vert_{L^2}$ is bounded 
(according to the $L^2$-norm of $\zeta$) 
and $\zeta(0)=\zeta_{1}(0)$ we deduce the 
existence of a constant $C_0$ such that for any $\lambda \leq 1$,
$$
{1 \over \lambda}
\Vert \psi_2(z+\lambda\zeta_{1})-\psi_2(z+\lambda\zeta) \Vert_{L^2}
\leq C_{0} \Vert \zeta_{1} - \zeta \Vert_{H^{1,2}}
\eqno(4.12)
$$
Now, by the Lebesgue Convergence Theorem and Remark 4.4 it is not
difficult to see that
the linear operator $d\psi_{2}(z)$ is continuous as linear map from 
$T_{z} \P$ to $L^{2}([0,1],\r)$. Then for any $\ep > 0$ there exists
$\vartheta \in ]0,\ep[$ such that 
$$
\Vert \zeta_{1} - \zeta \Vert_{H^{1,2}}
\leq \vartheta
\Longrightarrow
\Vert d\psi_2(z)[\zeta]-d\psi_2(z)[\zeta_{1}] \Vert_{L^2}<\ep.
\eqno(4.13)
$$
Moreover always fixing $\zeta_{1}$ such that
$\Vert \zeta_{1} - \zeta \Vert_{H^{1,2}}
\leq \vartheta$, for any $\lambda \leq 1$ we have 
$$
{1 \over \lambda}
\Vert \psi_2(z+\lambda\zeta_{1})-\psi_2(z+\lambda\zeta) \Vert_{L^2}
\leq C_{0} \vartheta \leq C_{0} \ep.
\eqno(4.14)
$$
Finally 
$$\displaylines{
{
{\psi_2(z+\lambda\zeta)-\psi_2(z)}
\over
{\lambda} 
}
-d\psi_{2}(z)[\zeta] = \cr
{
{\psi_2(z+\lambda\zeta_{1})-\psi_2(z)}
\over
{\lambda} 
}
-d\psi_{2}(z)[\zeta_{1}] +
{
{\psi_2(z+\lambda\zeta)-\psi_2(z+\lambda\zeta_{1})}
\over
{\lambda}
} 
+
d\psi_{2}(z)[\zeta_{1}]
-d\psi_{2}(z)[\zeta]. \cr }
$$
Since $\psi_{2}$ is differentiable at $z$ along the direction $\zeta_{1}$,
for any $\lambda$ sufficiently small it is
$$
\left\Vert
{
{\psi_2(z+\lambda\zeta_{1})-\psi_2(z)}
\over
{\lambda}
}
-d\psi_{2}(z)[\zeta_{1}]
\right\Vert_{L^2}
\leq \ep,
$$
so, combining (4.13) and (4.14) 
gives the existence of 
$\hat \lambda$ such that $|\lambda| \leq \hat \lambda$ implies
$$
\left\Vert
{
{\psi_2(z+\lambda\zeta)-\psi_2(z)}
\over
{\lambda}
}
-d\psi_{2}(z)[\zeta]
\right\Vert_{L^2}
\leq \ep + C_{0} \ep +\ep,
$$
proving that (4.11) is satisfied for any $\zeta \in H^{1,2}$.

The continuity of $d\psi_{2}(\cdot)$ 
(which is a consequence of Remark 4.4 and Lebesgue Theorem) 
says that $\psi_{2}$ is of class $C^1$ in $\P$ and its 
differential is given by (4.11).
\cvd
\2
\noindent
{\bf Remark 4.6.} 
If $z \in \Q$ then $\dot z \not= 0$ almost everywhere and it is lightlike.
Since 
$\langle Y(z),\dot z \rangle$ is negative 
and $Y$ is time like, there exists a positive constant $\nu_{z}$ such that 
$$
- \langle Y(z),\dot z \rangle \geq 
\nu_{z} \Vert Y(z) \Vert_{R} \Vert \dot z \Vert_{R}.
$$
Moreover $\langle Y(z),\dot z \rangle$ is constant, therefore
$\Vert \dot z \Vert_{R}$ is uniformly bounded.

\2
\noindent
Now we can finally prove that $\Q$ is a manifold in a neighborhood of $w$.

\2
\noindent
{\bf Proposition 4.7.} 
{\it There exists an open neighborhood ${\cal O}_{w}$ of the 
geodesic $w$ in $\Om$ such that $\Q \cap {\cal O}_{w}$ is 
a manifold of class $C^1$ and, for any 
$z \in \Q \cap {\cal O}_{w}$,}
$$\displaylines{
T_{z}{(\Q \cap {\cal O}_{w})} = 
\Big\{ \zeta \in \Om : \langle D_{s}{\zeta},\dot z
\rangle = 0 
\hbox { a.e., and}\cr\qquad
\langle Y,D_{s}\zeta \rangle + \langle D_{\zeta}Y,\dot z \rangle = 
\int_{0}^{1}({\langle Y,D_{s}\zeta \rangle +
\langle D_{\zeta}Y,\dot z \rangle})\;{\rm d}s \hbox { a.e., }  \Big\}\cr}
$$

\noindent
{\bf Proof.}  
Let $\psi : \P \longrightarrow L^{2}([0,1]\r)$ be the 
$C^1$-map  given by (4.9). 
By (4.10) its differential at any point $z \in \Q$ is
$$
d\psi(z)[\zeta] = 
{
{\langle D_{s} \zeta, \dot z \rangle}
\over
{\sqrt{\langle \dot z,\dot z \rangle_{R}}}
}.
$$
Since our result is of local nature 
(in a neighborhood of the geodesic $w$), by (4.7) it suffices 
to show that, for any 
$h \in  L^{2}([0,1]\r)$ , there exists $\zeta \in T_{z}\Om$ such that 
$$
\cases{ 
\langle Y(w),D_{s}\zeta \rangle + \langle D_{\zeta}Y(w),\dot w \rangle = 
\int_{0}^{1}({\langle Y,D_{s}\zeta \rangle +
\langle D_{\zeta}Y(w),\dot w \rangle})ds \cr
\displaystyle{
{
{\langle D_{s}\zeta, \dot w \rangle}
\over
{\sqrt{\langle \dot w,\dot w \rangle_{R}}}
}
} = h. \cr}
$$
Choose $\zeta(s) = \mu(s)Y(w(s)) + \lambda(s) \dot w(s)$.
Since $\langle Y, Y \rangle \equiv -1$, denoting by $c_w$ the real constant
$\langle Y,\dot w \rangle$, it will be sufficient
to verify the existence of  two real functions
$\mu$ and $\lambda$ such that $\mu(0)=0$, $\lambda(0)=\lambda(1)=0$ and
$$
\cases{ 
\mu' + \mu \langle D_{Y}Y,\dot w \rangle + \lambda' c_{w} & is constant, 
\cr
\mu' c_{w} = h{\sqrt{\langle \dot w,\dot w \rangle_{R}}}, \cr}
$$
and this can be done choosing
$$
\mu(s)=
{
{1} \over
{c_{w}}
}
\int_{0}^{s}{h{\sqrt{\langle \dot w,\dot w \rangle_{R}}}dr},
$$
and
$$
c_{w} \lambda' =
c + 
{
{1} \over
{c_{w}}
}
h{\sqrt{\langle \dot w,\dot w \rangle_{R}}} -
{
{\langle D_{Y}Y,\dot w \rangle}
\over
{c_{w}}
}
\int_{0}^{s}{h{\sqrt{\langle \dot w,\dot w \rangle_{R}}}dr},
$$
where (integrating both terms of the above equality) 
the constant $c$ can be choose so that 
$\lambda(1)=\lambda(0)=0$.
\cvd
\2
\noindent
{\bf Remark 4.8.} 
To describe the tangent space $T_{z}(\Q \cap {\cal O}_{w})$ 
we can operate in the following way. 
Take $\zeta \in T_{z}\P$ and choose 
$\mu \in H^{1,2}([0,1],\r)$ such that $\mu(0)=0$ and 
$$\langle D_{\dot z}[\zeta - \mu Y],\dot z \rangle = 0,$$
namely, $\mu$ has to satisfy
$$
\cases{ 
\langle D_{\dot z}\zeta,\dot z \rangle -
\mu' c_{z} - \mu \langle D_{\dot z}Y,\dot z \rangle = 0, 
\cr
\mu(0)=0, \cr}
$$
where $c_z \equiv \langle Y,\dot z \rangle$ is a negative constant.
Then $\mu$ is given by
$$
\mu(s)=
\int_{0}^{s}
{
{\langle D_{\dot z}\zeta,\dot z \rangle}
\over
{c_{z}}
}
exp(-\int_{r}^{s}
{
{\langle D_{\dot z}Y,\dot z \rangle}
\over
{c_{z}}
}d\sigma)dr,
$$
and, by (4.16),
$$
d\tau(z)[\zeta-\mu Y] =
-\langle \dot \gamma (\tau(z),\zeta(1) \rangle +
\langle \dot \gamma(\tau(z)),Y(z(1) \rangle \mu(1)
$$
where 
$$
\mu(1) = 
\int_{0}^{1}
{
{\langle D_{\dot z}\zeta,\dot z \rangle}
\over
{c_{z}}
}
exp(-\int_{r}^{1}
{
{\langle D_{\dot z}Y,\dot z \rangle}
\over
{c_{z}}
}d\sigma)dr.
$$
Then, using Remark 4.4 and the same technique of the proof of Lemma 4.5
allows to deduce that $\tau$ is of class $C^2$ on $\Q \cap {\cal O}_{w}$.

\1
In sections 2 we shows the existence 
(for any timelike curve sufficiently closed to a fixed event) 
of  minimizing lightlike geodesics for the functional $\tau$. 
Now we need a sort of converse of the above principle.

\2
\noindent
{\bf Proposition 4.9.} 
{\it Any future pointing lightlike geodesic $w$ is a 
critical point of $\tau$ on $\Q \cap {\cal O}_{w}$.}

\1

\noindent
{\bf Proof.} 
Take $V \in T_{w}\Om$ such that $V(0)=V(1)=0$ and  
$\mu \in H^{1,2}([0,1],\r)$ such that $\mu(0)=0$  and
$$
\langle D_{\dot w}[V-\mu Y], \dot w \rangle = 0.
\eqno(4.17)
$$
By (4.16), Proposition 4.7 and Remark 4.8 it will be sufficient to prove that 
$$
0 = d\tau(w)[V - \mu Y] = \mu(1) \langle Y(w(1)),\dot \gamma(\tau(w)) \rangle
\eqno(4.18)
$$
Now by (4.17) 
$$
\mu(1) = 
\int_{0}^{1} 
{
{\langle D_{\dot w}V,\dot w \rangle}
\over
{\langle Y,\dot w \rangle}
}ds
$$
because $D_{\dot w} = 0$. Then $\mu(1)=0$ for any $V$ because 
$\langle Y,\dot w \rangle$ is constant ($\not=0$) and $w$ is a geodesic.
\cvd
\2
\noindent
{\bf Remark 4.10.} The above proof shows also that, for any 
$\zeta \in T_{w}(\Q \cap {\cal O}_{w}$,
since $w$ is a geodesic it is $\zeta(1) = 0$.

\2
\noindent
{\bf Remark 4.11.} Let $w$ be a future pointing lightlike geodesic.
The same computations in [22] allows to prove that, for any
$\zeta \in T_{w} \Q$, the Hessian of $\tau$ along the direction $\zeta$
is given by
$$
H^{\tau}(w)[\zeta,\zeta] = 
{
{-1}
\over
{\langle \dot \gamma(\tau(w)),\dot w(1) \rangle}
}
\int_{0}^{1} ({\langle D_{s} \zeta, D_{s} \zeta \rangle -
\langle R(\zeta,\dot w)\dot w,\zeta} \rangle) ds.
\eqno(4.19)$$

\1
\noindent
Now we equip $T_w \Q$ with the Hilbert structure
$$
\langle \zeta_1,\zeta_2 \rangle = \int_{0}^{1} \langle D_{s}^{Y}\zeta_1,
D_{s}^{Y}\zeta_2 \rangle_{Y} ds,
\eqno(4.20)
$$
where $\langle \zeta,\zeta \rangle_Y = \langle \zeta,\zeta \rangle - 2 \langle
Y(w),\zeta \rangle^2$ (which is equivalent to (2.1)) and $D_{s}^{Y}$
is the covariant derivative with respect to $\langle \zeta,\zeta \rangle_Y$.

\2
\noindent
{\bf Proposition 4.12.} 
{\it The linear map associated to the quadratic form (4.19) on $T_w \Q$ 
is a compact perturbation of the identity with respect to the 
Hilbert structure (4.20)}.

\1
\noindent
{\bf Proof.} It is 
$$
\displaylines{
\langle D_{s}\zeta,D_{s}\zeta \rangle =
\langle D_{s}\zeta,D_{s}\zeta \rangle +
2\langle D_{s}\zeta,Y(w) \rangle^{2} -
2\langle D_{s}\zeta,Y(w)\rangle^{2} =  \cr
\hfill \langle D_{s}\zeta,D_{s}\zeta \rangle_{Y} -2\langle
D_{s}\zeta,Y(w)\rangle^{2}.
\hfill \llap(4.21) \cr}
$$
Now there exists a bilinear map $\Gamma$ defined on the vector 
fields on $\Lambda$ such that 
$$
D_{s}\zeta = D_{s}^{Y}\zeta + \Gamma(w)[\zeta,\zeta].
$$
Moreover, by (4.15)
$$ 
\langle D_{s}\zeta,Y(w) \rangle = -\langle D_{\zeta}Y,\dot w\rangle 
+\int_{0}^{1}
({\langle D_{s}\zeta,Y(w) \rangle + \langle D_{\zeta}Y,\dot w\rangle}) ds
\eqno(4.22)
$$
while, by Remark 4.10
$$
\int_{0}^{1}
{\langle D_{s}\zeta,Y(w) \rangle} = 
-\int_{0}^{1}{\langle \zeta,D_{s}Y \rangle}.
\eqno(4.23)
$$
Since $H^{1,2}$ is compactly embedded in $L^\infty$, combining 
(4.21), (4.22) and (4.23) gives the proof.
\cvd
\1      
Denote now by $m(w,\tau)$ the maximal dimension of a subspace of 
$T_{w}(\Q \cap {\cal O}_{w})$ where the restriction of  
$H^{\tau}(w)[\cdot,\cdot]$ is negative definite. 
It is called the Morse Index of the quadratic form 
$H^{\tau}(w)[\cdot,\cdot]$. The following Index Theorem holds:

\2
\noindent
{\bf Theorem 4.13.} {\it Let $w$ be a geodesic in $\Q$. Then}
$$
m(w,\tau) = \mu(w).
$$

\1
\noindent
To prove Theorem 4.13 some preliminary results are needed.

\2

\noindent
{\bf Lemma 4.14.} {\it Let $\zeta$ be a Jacobi field along $w$ such that 
$\zeta(0)=0, \zeta(1)=0$. Then} $\zeta \in T_w\Q$.
\1
\noindent
{\bf Proof.} 
Let $\zeta$ be a Jacobi field along $w$ with $\zeta(0)=0$ and
$\zeta(1)=0$.  
It is immediately checked that,$\langle \dot w,D_{s}\zeta \rangle \equiv 0.$
Therefore we have just to prove that the function 
$$
\varphi(s)=\langle D_{s}\zeta,Y(w) \rangle + \langle \zeta,D_{\zeta}Y \rangle
$$
is constant. Since $D_{s}Y \equiv 0$ and $w$ is a geodesic
$$
\varphi'(s)=
\langle D_{s}^{2}\zeta,Y(w) \rangle + \langle \zeta,D_{s}D_{\zeta}Y \rangle.
$$
Now, since $\zeta$ is a Jacobi field and 
$$
D_{s}D_{\zeta}Y = D_{\zeta}D_{s}Y + R(\dot w,\zeta)Y
$$
(cf [2]), it is 
$$
\displaylines{ 
\varphi'(s)=
-\langle R(\zeta,\dot w)\dot w, Y(w) \rangle +
\langle D_{\zeta}D_{s}Y + R(\dot w,\zeta)Y,\dot w \rangle = \cr
-\langle R(\zeta,\dot w)\dot w, Y(w) \rangle +
\langle R(\dot w,\zeta)Y,\dot w \rangle = 0 \cr}
$$
because of the symmetry properties of $R$ (cf. [2]). 
\cvd
\2
\noindent
An integration by parts shows immediately that the following Lemma holds.

\2

\noindent
{\bf Lemma 4.15.} 
{\it If $\zeta$ is a Jacobi field along $w$ such that $\zeta(0) = 0$ and 
$\zeta(1) = 0$, then}
$$
H^{\tau}(w)[\zeta,\zeta_1] = 0 \hbox{ for any } \zeta_1 \in T_{w} \Q
\eqno(4.24)
$$
\2
\noindent
{\bf Lemma 4.16.} {\it Let $\zeta \in T_w \Q$ such that {\rm (4.24)} is
satisfied. Then $\zeta$ is a $C^2$-Jacobi 
field along $w$ such that $\zeta(0) = 0$ and $\zeta(1) = 0$.}

\1
\noindent
{\bf Proof.} 
If $\zeta \in T_{w}\Q$, $\zeta(0)=0$ and by Remark 4.9, 
$\zeta(1)=0$ because 
$w$ is a geodesic. 
Then, assuming that  (4.24) holds, 
we have to verify that $\zeta$ is of class $C^2$ and it satisfies
(1.4).

Let $V$ be a $C^\infty$-vector field along $w$ such that $V(0)=0,\
V(1)=0$.  Set $c_w = \langle Y(w), \dot w \rangle$ 
(which is a non zero constant) and 
choose
$$
\mu(s) = \int_{0}^{s} {{1}\over{c_w}} \langle D_{s}V,\dot w \rangle.
$$
Now let $\lambda$ be the unique real map such that 
$$
\displaylines{
\langle D_{s}V,Y(w) \rangle + 
\langle \dot w, D_{V}Y \rangle -\mu' \langle Y(w),Y(w) \rangle
-\mu \langle \dot w, D_{Y}Y \rangle -\lambda' c_w = const \cr
\lambda(0) = \lambda (1) = 0. \cr}
$$ 
A straightforward computation shows that 
$\zeta_{1} = V - \mu Y(w) -\lambda \dot w \in T_w \Q$.
Therefore by (4.24) we have
$$
\int_{0}^{1} \langle D_{s}\zeta, D_{s}V-\mu' Y(w)-\lambda' \dot w \rangle
- \langle R(\zeta,\dot w)\dot w, V-\mu Y(w) - \lambda \dot w \rangle = 0
$$
for any $C_{0}^{\infty}$-vector field $V$ along $w$.
Now $\langle D_{s}\zeta, \dot w \rangle \equiv 0$ and 
$\langle R(\zeta,\dot w)\dot w, \dot w \rangle \equiv 0$.
This allows us to deduce immediately that $\zeta$ is of class $C^2$. So, 
to obtain (1.4), 
it suffices to prove that, for any $V$,
$$
\int_{0}^{1} -\mu' \langle D_{s}\zeta,Y(w)\rangle
+\mu \langle R(\zeta,\dot w)\dot w,Y(w) \rangle = 0
\eqno(4.25)
$$
which is equivalent to
$$
\int_{0}^{1} -\langle D_{s}V,\dot w \rangle \langle D_{s}\zeta,Y(w) \rangle
+\langle V, \dot w \rangle \langle R(\zeta,\dot w)\dot w,Y(w) \rangle = 0
$$
Then an integration by parts shows that (4.25) is equivalent to
$$
{
{d}\over{ds}
}
\left(
\langle D_{s}\zeta,Y(w) \rangle \dot w) + 
\langle R(\zeta,\dot w)\dot w,Y(w) \rangle
\right)
= 0. 
$$
Now, $w$ is a geodesic while, by (4.15),
$$ 
\langle D_{s}\zeta,Y(w) \rangle = \left( \int_{0}^{1}
\langle D_{s}\zeta,Y(w) \rangle + \langle D_{\zeta}Y, \dot w \rangle \right)
-\langle D_{\zeta}Y, \dot w \rangle
$$
therefore (4.25) is equivalent to
$$
-{
{d}\over{ds}
}
(\langle D_{\zeta}Y,\dot w \rangle) + 
\langle R(\zeta,\dot w)\dot w,Y(w) \rangle = 0.
$$
But 
$$
D_{s}D_{\zeta}Y = D_{\zeta}D_{s}Y + R(\dot w,\zeta)Y,
$$
$w$ is a geodesic and $D_{s}Y \equiv 0$, therefore (4.25)
follows by the symmetry properties of $R$.
\cvd
\2
\noindent
{\bf Proof of Theorem 4.13.} 
We prove a generalization of the Morse Index Theorem for Riemannian
geodesics (cf. for instance [16,17]) to lightlike geodesics.
For any $\sigma \in ]0,1]$ set 
$$\displaylines{
T_{\sigma} = \Big\{ \zeta \in H^{1,2}([0,\sigma]), {\cal T} {\Lambda}):
{\zeta}(s) \in {T_{w(s)}} \hbox{ for any } s \in ]0,\sigma], \cr
\zeta(0) = 0, \zeta(\sigma) = 0, \cr 
\langle Y,D_{s}{\zeta} \rangle + \langle D_{\zeta}{Y},\dot w \rangle \equiv
c_{\zeta} \hbox { a.e. in } [0,\sigma] \cr
\langle D_{s}{\zeta}, \dot w \rangle = 0 \hbox { a.e. in } [0,\sigma] \}. \cr}
$$
Note that $T_1 = T_{w} \Q$. Set, for any $\zeta \in T_{\sigma}$ 
$$
H_{\sigma}([\zeta,\zeta] = 
\int_{0}^{\sigma} ({\langle D_{s} \zeta, D_{s} \zeta \rangle -
\langle R(\zeta,\dot w)\dot w,\zeta} \rangle) ds.
$$
Note that 
$$
H^{\tau}(w)[\cdot,\cdot] = 
{
{-1}
\over
{\langle \dot \gamma(\tau(w)),\dot w(1) \rangle}
}
H_{1}([\cdot,\cdot],
$$
while $-\langle \dot \gamma(\tau(w)),\dot w(1) \rangle >0$ because
$$\dot \gamma(\tau(w)) = Y(\gamma(\tau(w))) = Y(w(1))$$ and 
$\langle Y(w(s)),\dot w(s) \rangle$ is constant.

Therefore, denoting by $i(H_1)$ 
the Morse Index of the quadratic form $H_1$ we have to prove that 
$$
i(H_1) = \mu(w)
\eqno(4.26)
$$
On $T_{\sigma}$ we define the Hilbert structure
$$
\langle \zeta,\zeta \rangle_{\sigma} = 
\int_{0}^{\sigma} \langle D_{r}^{Y}\zeta,D_{r}^{Y}\zeta \rangle_{Y} dr.
\eqno(4.27)
$$
Denote by $L_{\sigma}$ the linear operator in $T_{\sigma}$ such that
$$
\langle L_{\sigma}\zeta_1,\zeta_2\rangle_{\sigma} = 
H_{\sigma}([\zeta_1,\zeta_2]).
\eqno(4.28)
$$
By Proposition 4.12 we see that 
$$
L_{\sigma} = I_{\sigma} - K_{\sigma}
$$
where $I_{\sigma}$ is the identity on 
$T_{\sigma}$ and $K_{\sigma} : T_{\sigma} \rightarrow T_{\sigma}$ 
is a compact operator.
We shall denote by $w_\sigma$ the geodesic 
$w_{|[0,s]}$. 
It is well known that $T_{\sigma}$ has the 
following orthogonal decomposition consisting of eigenspaces of $L_\sigma$
$$
T_\sigma = H_{\sigma}^{+} \oplus H_{\sigma}^{0} \oplus  H_{\sigma}^{-} 
$$
where (4.28) is positive definite on  $H_{\sigma}^{+}$, definite 
negative on $H_{\sigma}^{-}$
and  $H_{\sigma}^{0} = Ker L_{\sigma}$.

For any $\sigma \in ]0,1]$ let 
$\lambda_{1}(\sigma) \geq ... \geq \lambda_{k}(\sigma)$ 
be the eigenvalues of the compact operator $K_{\sigma}$. 
Any eigenvalues is here repeated according with 
its (finite) multiplicity. 
By Lemmas 4.15-4.17 $w(\sigma)$ is conjugate 
to $w(0)$ with multiplicity equal to $m$ if and only if 
there exists $k \in \n$ such that 
$$\lambda_k(\sigma)<1,\quad
\lambda_{k+1}(\sigma),...,\lambda_{k+m}(\sigma) = 1,\quad{\rm and}\quad
\lambda_{k+m+1}(\sigma)>1.
$$
Since $K_{\sigma}$ is compact the eigenvalues of 
$K_{\sigma}$ are characterized by the well known Poincar\'e Formula:
$$
\lambda_{k}(\sigma) = \displaystyle \max_{\hbox{dim }V=k} \Big[
\displaystyle \min_{\zeta \in V,  |\zeta|_{\sigma}=1} 
(\langle K_{\sigma} \zeta,\zeta \rangle _{\sigma})\Big]
\eqno(4.29)
$$
where $|\cdot|_{\sigma} = \langle \cdot,\cdot \rangle_{\sigma}^{1/2}$.
Using (4.29) it is not difficult to prove that any $\lambda_k$ is
a continuous function of $\sigma$. 
Moreover using (4.21)-(4.23) it is easy to show that there
exists $\sigma_{0} > 0$ such that, for any $\sigma \in ]0,\sigma_{0}]$,
$H_{\sigma}$ is positive definite on $T_{\sigma}$. 
Then it  will be sufficient to show that
$$
\hbox{ any $\lambda_{k}$ is strictly increasing.}
\eqno(4.30)
$$
Fix $k \in \n$ and $0 \leq \sigma_1 < \sigma_2 \leq 1$.
By (4.29) there exists a subspace $V$ of 
$T_{\sigma_{1}}$ having dimension $k$ such that
$$
\lambda_{k}(\sigma_{1}) =  
\displaystyle\min_{\zeta \in V, \die |\zeta|_{\sigma_{1}}=1}
(\langle K_{\sigma_{1}} \zeta,\zeta \rangle _{\sigma_{1}}).
\eqno(4.31)
$$
For any 
$\zeta \in V$ 
set 
$c_{\zeta} \equiv 
\langle D_{s}\zeta,Y(w) \rangle + \langle \dot w,D_{\zeta}Y \rangle$ and 
$c_{w} \equiv \langle \dot w,Y(w) \rangle$. Take the vector field
$$A_{\zeta} = \zeta + \lambda \dot w$$
where $\lambda' = {
{-c_{\zeta}} \over {c_{w}}
}$ and $\lambda (\sigma_1)=0$. Since $\zeta \in T_{\sigma_{1}}$,
$w$ is a lightlike geodesic
and $D_{s}Y \equiv 0$ it is 
$$
\langle D_{s}A_{\zeta},\dot w \rangle \equiv 0,
\eqno(4.32)
$$
$$
\langle D_{s}A_{\zeta},Y(w) \rangle + 
\langle \dot w, D_{A_{\zeta}}Y \rangle \equiv 0,
\eqno(4.33)
$$
and
$$
A_{\zeta}(\sigma_1) = 0, \die A_{\zeta}(0) = \lambda(0) \dot w(0).
\eqno(4.34)
$$
Now denote by 
$\hat{A_{\zeta}}$ the extension to 
$[0,\sigma_2]$ of $A_{\sigma}$ obtained by setting 
$\hat {A_{\zeta}} = 0$ on $[\sigma_1,\sigma_2]$. 
Note that, for any $\zeta \in V$
$$
\langle K_{\sigma_{2}} \hat{A_{\zeta}},\hat{A_{\zeta}} \rangle_{\sigma_{2}} =
\langle K_{\sigma_{1}} {\zeta},{\zeta} \rangle_{\sigma_{1}}.
$$
Now set
$$
B_{\zeta} = \hat {A_{\zeta}} + \mu \dot w
$$
where $\mu$ satisfies
$$
\mu' = \hbox{const }, \mu(0)=-\lambda(0), \die \mu(\sigma_{2})=0.
\eqno(4.35)
$$
Since $\hat {A_{\zeta}}$ satisfies (4.32)-(4.34) (with $A_\zeta$ replaced by
$\hat{A_{\zeta}}$), thanks to (4.35) we deduce that 
$B_{\zeta} \in T_{\sigma_{2}}$.
Note that the map
$$B_{\cdot} : V \longrightarrow T_{\sigma_{2}}$$
is a linear and injective.
Then the space 
$$
V_{*} = \Big\{B_{\zeta}: \zeta \in V \}
$$
is a subspace of $T_{\sigma_{2}}$ having dimension $k$.
Moreover, by our construction,
$$
\langle K_{\sigma_{2}} B_{\zeta},B_{\zeta} \rangle_{\sigma_{2}} =
\langle K_{\sigma_{1}} {\zeta},{\zeta} \rangle_{\sigma_{1}}
\hbox{ for any } \zeta \in V.
$$
Then, by (4.29) and (4.31)
$$
\displaylines{
{\lambda_{k}}({\sigma_{1}}) =  
\displaystyle \min_{{\zeta} \in V, |{\zeta}|_{\sigma_{1}}=1}
( \langle K_{\sigma_{1}} \zeta,\zeta \rangle _{\sigma_{1}} ) = \cr
\hfill
\displaystyle \min_{{\zeta} \in V_{*}, |{\zeta}|_{\sigma_{2}}=1}
( \langle K_{\sigma_{2}} \zeta,\zeta \rangle_{\sigma_{2}} )
\leq {\lambda_{k}}(\sigma_2).
\hfill \llap(4.36) \cr}
$$
To conclude the proof assume by contradiction that 
$$
\lambda \equiv \lambda_{k}(\sigma_{1}) = \lambda_{k}(\sigma_{2}).
$$
By the spectral properties of $K_{\sigma_{2}}$, $T_{\sigma_{2}}$
admits the orthogonal decomposition
$$
T_{\sigma_{2}} = H^{-} \oplus H^{0} \oplus H^{+},
$$
such that $\lambda I_{\sigma_{2}} - K_{\sigma_{2}}$
is negative definite on $H^{-}$, positive definite on $H^{+}$
and $H^{0} = Ker(\lambda I_{\sigma_{2}} - K_{\sigma_{2}})$.

We claim that 
$$
V_{*} \cap ( H^{0} \oplus H^{+} ) = \{ 0 \}.
\eqno(4.38)
$$
Indeed if $\zeta = {\zeta}_{0} + {\zeta}_{+} \in 
V_{*} \cap ( H^{0} \oplus H^{+} )$ , where ${\zeta}_{0} \in H^{0}$ and
${\zeta}_{+} \in H^{+}$,
if ${\zeta}_{+} \not= 0$ and $|{\zeta}|_{\sigma_{2}} = 1$ it is
$$
\displaylines{
\langle K_{\sigma_{2}} {\zeta}_{0},{\zeta}_{0} \rangle_{\sigma_{2}}
+ \langle K_{\sigma_{2}} {\zeta}_{+},{\zeta}_{+} \rangle_{\sigma_{2}} =
\lambda \langle {\zeta}_{0},{\zeta}_{0} \rangle_{\sigma_{2}}
+ \langle K_{\sigma_{2}} {\zeta}_{+},{\zeta}_{+} \rangle_{\sigma_{2}} <
\cr
\lambda \langle {\zeta}_{0},{\zeta}_{0} \rangle_{\sigma_{2}}
+ \lambda \langle {\zeta}_{+},{\zeta}_{+} \rangle_{\sigma_{2}} = \lambda
\cr }
$$
in contradiction with (4.36) and (4.37) because
${\zeta} \in V_{*}$.
Then ${\zeta}_{+} = 0$ and
$$
{\zeta} = {\zeta}_{0} \in Ker( \lambda I_{\sigma_{2}} - K_{\sigma_{2}}).
$$
Then the same proof of Lemma 4.16 allows to deduce that $\zeta$ is of class
$C^2$. Since $\mu \dot w$ is of class $C^2$ then $\hat{A_{\zeta}}$
is of class $C^2$, and by the construction of  $\hat{A_{\zeta}}$
we deduce that 
$$
D_{s}(A_{\zeta})({\sigma}_{1}) = 0.
$$
Since $\zeta$ is a Jacobi field in $[0,{\sigma}_{1}]$, $w$ is a geodesic
and $\lambda'$ is constant, $A_{\zeta} = \zeta + \lambda \dot w$
satisfies (1.4) with initial condition $A_{\zeta}({\sigma}_{1})=0$ and
$D_{s} (A_{\zeta}) ({\sigma}_{1}) = 0$. Then by the uniqueness of
the Cauchy problem it is $A_{\zeta} \equiv 0$. Since ${\zeta}(0)=0$, then
${\lambda}(0)=0$
and therefore $\lambda \equiv 0$. This imply that $\mu \equiv 0$ and
$B_{\sigma} \equiv 0$ proving (4.38).
Then 
the orthogonal projection of
$V_{*}$ on $H^{-}$ has dimension $n$ and 
$$
{\lambda}_{k} ({\sigma}_{1}) = \lambda <
\displaystyle \min_{ {\zeta} \in V_{*},|{\zeta}|_{\sigma_{2}}=1}
\langle K_{\sigma_{
2}} {\zeta},{\zeta} \rangle_{\sigma_{2}})
\leq {\lambda}_{k} ({\sigma}_{2})
$$
proving (4.30) and concluding the proof of Theorem 4.13.
\cvd

\2
\centerline{\bf 5. PROOFS OF THEOREMS 1.5, 1.9 AND 1.11}\vs
\1
\noindent
Now we are finally ready to prove Theorem 1.5.

\2
\noindent 
{\bf Proof of Theorem 1.5.}  Whenever we are far from the geodesics in 
${\cal L}^{+,2}_{p,\gamma}(\Lambda)$, 
we can use the shortening flow at section 3 
to obtain a flow where $\tau$ is strictly decreasing. 
Nearby any geodesic we can construct an 
homotopy equivalence between $\L$ and $\Q$ 
simply by a convex combination between the identity in 
$H^{1,2}([0,1],\r)$ and the parameterization described by the condition
$$
\langle \dot z, Y(z) \rangle = 
\int_{0}^{1}{\langle \dot z, Y(z) \rangle}ds \hbox { a.e. }
$$
Then, we can use the shortening flow for 
$\tau$ far from geodesics and, thanks to Proposition 4.7, Remark 4.8, 
and Proposition 4.9, we can use the classical Morse Theory
(cf. e.g. [4,17]) to describe the topology nearby a geodesic. 
In this way we obtain
$$
\sum_{w\in {\cal G}_{p,\gamma}(\Lambda)} \lambda^{m(w,\tau)}=
{\cal P}({\cal L}_{p,\gamma}^{+,2}(\Lambda))(\kappa)+
(1+\kappa)S(\kappa)
$$
where $S$ is a formal series  with coefficients in $\n \cup\{+\infty\}$.

Since ${\cal B}^+_{p,\gamma}(\Lambda)$ is homotopically equivalent
to $\L$ (cf. Proposition 3.4), applying Theorem 4.13
gives the proof of (1.6).
\cvd

\2
\noindent
{\bf Proof of Theorem 1.9.} Assume that ${\cal B}^+_{p,\gamma}(\Lambda)$ is
contractible. 
Then the Poincar\'e polynomial of ${\cal B}^+_{p,\gamma}({\Lambda})$
with respect to any field $\k$ is given by
$${\cal P}({\cal B}^{+}_{p,\gamma}(\Lambda),\k)(\kappa)=1$$
Let ${\cal G}^+_{p,\gamma}({\Lambda})$ 
be the set of future pointing lightlike
geodesics joining $p$ with $\gamma$. Formula (1.6) gives
$${\rm card}{\cal G}^{+}_{p,\gamma}({\Lambda})=1+2 S(1).$$
Then ${\rm card}{\cal G}^{+}_{p,\gamma}(\Lambda)$ is odd or infinite, 
according if $S(1)$ is finite or infinite. 
If ${\cal B}^+_{p,\gamma}(\Lambda)$ is not contractible,
cat$\, {\cal B}^+_{p,\gamma}(\Lambda)\ge 2$ 
so the conclusion follows by Remark 3.9.\cvd

\non
{\bf Proof of Theorem 1.11.} If $\Lambda$ is contractible, then
$\Omega(\Lambda)$ is
contractible. Then by assumption $L_3$), ${\cal B}^+_{p,\gamma}(\Lambda)$
is contractible and the proof follows by the first part of Theorem 1.9.

If $\Lambda$ is not
contractible by $L_3$) and a result of [29], for a suitable $\k$
${\cal P}({\cal B}^+_{p,\gamma}(\Lambda))(\kappa)$ 
has infinitely many coefficients
different from zero and the conclusion follows by formula (1.6).\cvd

\2
\centerline
{\bf APPENDIX A.\ ON THE TOPOLOGY OF ${\cal B}^+_{p,\gamma}(\Lambda)$.}

\1
Under light--convexity and pseudo--coercivity of $\tau$ we have seen
(in section 3) that ${\cal B}^+_{p,\gamma}(\Lambda)$ is homotopically
equivalent to ${\cal L}^{+,r}_{p,\gamma}(\Lambda)$ (for any $r\in
[1,+\infty]$).

In this appendix we shall give a general condition assuring that
${\cal L}^{+,1}_{p,\gamma}(\Lambda)$ is homeomorphic to
$\Omega^{1,1}_{p,\gamma}(\Lambda)$. Then, using standard technique
one see that $\Omega^{1,1}_{p,\gamma}(\Lambda)$ is homotopically equivalent
to the based loop space of $\Lambda$.
\non
{\bf Proposition A.1} {\it Suppose that there exists a smooth hypersurface
$\Lambda_0$ in $\Lambda$ and a smooth time--like vector field $Y$ in
$\Lambda$ such that

1) $Y$ is complete in $\Lambda$
2) $\Lambda=\{\eta(\sigma,y):y\in\Lambda_0,
\ \sigma\in\r,\ \dot\eta=Y(\eta),\
\eta(0)=y\}$
3) for any integral curve $\eta$ of $Y$ there exist a unique $s\in\r$ such
that $\eta(s)\in\Lambda_0$

4) $\gamma:\r\to\Lambda$ is an integral curve of $Y$ with
$\gamma(0)\in\Lambda_0$

5) $p\in\Lambda_0$ and $p\ne\gamma(0)$.
6) The Cauchy problem
$$ 
\cases{\sigma'={-1\over\langle Y,Y\rangle}\left( \langle Y,\eta_y[\dot
y]\rangle+{1\over 2} \sqrt{\langle Y,\eta_y[\dot
y]\rangle^2-\langle Y,
Y\rangle \langle \eta_y[\dot y],\eta_y[\dot y]\rangle}\right)\cr
\sigma(0)=0 \cr} 
\eqno(A.1)
$$
can be solved in the interval $[0,1]$ for any $y\in H^{1,1}([0,1],\Lambda)$
such that $y(0)=p$ and $y(1)=\gamma(0)$. 

Then ${\cal L}_{p,\gamma}^{+,1}(\Lambda)$ is homeomorphic to
$\Omega_{p,\gamma}^{1,1}(\Lambda)$.}

\non
(Here $Y=Y(\eta(\sigma,y)$) and
$\eta_y$ denotes the derivative of $\eta$ with respect to the second
variable).

\non
{\bf Proof.} Take $z(s)=\eta(\sigma(s),y(s))$. Suppose
$y \in H^{1,1}([0,1],\Lambda)$, $y(0)=p$ and $y(1)=\gamma(0).$ 
If $\sigma$ satisfies 6) a straightforward computation show 
that the curve $z(s)$ is in
${\cal L}^{+,1}_{p,\gamma}(\Lambda)$. Conversely if $z\in {\cal
L}^{+,1}_{p,\gamma}(\Lambda)$ can be projected on $\Lambda_0$ using the
integral curve of $Y$ (cf. assumption 3). Since (A.1) has a unique solution
we are done.\cvd

\2
\centerline {\bf APPENDIX B.\ MORSE RELATIONS ON THE SPACE}
\centerline {\bf OF THE PIECEWISE LIGHTLIKE GEODESICS.}

\1
In this appendix we show by a simple example that we can not write
Morse Relations using the topology of the piecewise (non null)
lightlike geodesics (endowed with the topology of the uniform convergence).
This space, as in section 1, will be denoted by
$\hat {\cal B}^{+}_{p,\gamma}(\Lambda)$.

On the space $\r^2 \times \r$ we consider the flat Minkowski metric
$$
ds^2 = dx_{1}^{2} + dx_2^{2} -dt^{2}.
$$
Take $p=(y_{0},0)$ and $\gamma(s) = (y_{1},s)$.
It is immediate to verify that 
$\hat {\cal B}^{+}_{p,\gamma}(\Lambda)$
is homeomorphic to the space ${\cal C}_{y_{0},y_{1}}$ of the piecewise
non null geodesics in $\r^{2}$ (with respect to the Euclidean metric)
joining $y_{0}$ with $y_{1}$ endowed with the uniform topology.
Considering the positions on the unit circle assumed by the 
unit speed of any broken geodesic
is not difficult to show that  ${\cal C}_{y_{0}.y_{1}}$ has infinitely
many connect components. Then, if Morse Relations
hold, one should obtain the existence of infinitely many geodesics
joining $p$ and $\gamma$ and this is clearly false.

Analogously one see that in the 
(2+1)-dimensional Minkowski spacetime the infinite 
dimensional space where the relativistic Fermat  Principle is proved, 
has infinitely many connect components. 
Then also in this case it is not possible to write the Morse Relations.

\vfil
\eject

\centerline{\bf REFERENCES}
\1
\item{[{}1]}
F. ANTONACCI, P. PICCIONE, A Fermat principle on Lorentzian
manifolds and applications, {\it Appl.Math.Letters}, {\bf 9} 91-96 (1996).
\item{[{}2] }
J.K. BEEM,  P. H. EHRLICH, K.L. EASLEY,
{\it Global Lorentzian Geometry}.
Marcel Dekker. New York, 1996.
\item{[{}3]  }
H. BREZIS,
{\it Analyse fonctionelle},
Masson: Paris, 1984.
\item{[{}4]  }
K.C. CHANG, 
{\it Infinite Dimensional Morse Theory and Multiple Solutions Problems}.
Birkhauser, Boston, 1993.
\item{ [{}5] }
V.  FARAONI, Nonstationary gravitational lenses and the Fermat
principle, {\it Astrophys. J.} {\bf 398}, 425--428 (1992).

\item{[{}6] }
D. FORTUNATO, F. GIANNONI,  A. MASIELLO,
A Fermat principle for stationary space-times
with applications to light rays,
{\it J. Geom. Phys.} {\bf 15}, 159-188 (1995).
\item{[{}7]}
A. GERMINARIO,
Morse Theory for light rays without nondegeneration assumptions,
{\it Nonlinear World} {\bf 4}, 173--206 (1997).
\item{[{}8]}
F. GIANNONI,  A. MASIELLO,
On a Fermat principle in General Relativity.
A Morse Theory for light rays,
{\it Gen. Rel. Grav.} {\bf 28}, 855-897 (1996).
\item{[{}9]}
F. GIANNONI, A. MASIELLO, P. PICCIONE,
A variational theory for light rays on stably causal Lorentzian
manifolds : regularity and multiplicity results,
{\it Comm. Math. Phys.} {\bf 187}, 375-415 (1997).
\item{[10]}
F. GIANNONI, A. MASIELLO,  P. PICCIONE,
A Morse theory for light rays on stably causal Lorentzian
manifolds, {\it Ann. Inst. H. Poincar\'e, Physique Theorique} {\bf 69},
359--412 (1998).
\item{[11]}
F. GIANNONI, A. MASIELLO, P. PICCIONE,
A  timelike extension of Fermat's
principle  in  General  Relativity  and  Applications,
{\it Calc. Var. and P.D.E.}, {\bf 6}, 263-283 (1998).
\item{[12]}
F. GIANNONI, A. MASIELLO and P.PICCIONE,
Convexity and the finiteness of the number of geodesics. 
Applications to the multiple image effect, 
{\it Class. Quantum Grav.} {\bf 69}, 731--748 (1999).
\item{[13]}
S.W. HAWKING,  G.F. ELLIS,
{\it The Large Scale Structure of Space-Time}.
Cambridge University Press, London/New York, 1973.
\item{[14]}
I. KOVNER
Fermat principles for arbitrary space-times,
{\it Astrophys. J.} {\bf 351}, 114-120 (1990).
\item{[15]}
T. LEVI-CIVITA,
{\it Fondamenti di Meccanica Relativistica}.
Zanichelli, Bologna 1928.
\item{ [16]}
A. MASIELLO, 
{\it Variational Methods in Lorentzian Geometry}.
Pitman Research Notes in Matematics {\bf 309}. 
Longman, London, 1994.
\item{[17]}
J. MAWHIN, M.WILLEM,
{\it Critical point theory and Hamiltonian systems}. Springer Verlag,
New York 1988.
\item{[18]}
J. MILNOR,
{\it Morse Theory}.
Princeton University Press, Princeton, 1963.
\item{[19]}
B. O'NEILL,
{\it   Semi-Riemannian   Geometry   with   applications  to
Relativity}.
Ac. Press, New-York-London, 1983.
\item {[20]}
R. PALAIS,
{\it Foundations of Global Nonlinear Analysis},
W.A.Benjamin, New-York, 1968.
\item {[21]}
V. PERLICK,
On Fermat's principle in General Relativity: I. The general case,
{\it Class. Quantum Grav.} {\bf 7}, 1319-1331 (1990).
\item{[22]}
V. PERLICK,
Infinite   dimensional   Morse   Theory   and   Fermat's   principle    in
general relativity. I,
{\it J. Math. Phys.} {\bf 36}, 6915-6928 (1995).
\item{[23]}
V. PERLICK, P.\ PICCIONE, 
A general-relativistic 
Fermat principle for
extended light sources and extended receivers,  
{\it Gen.\  Rel.\ Grav.\/} {\bf30}, n.\ 10, \ 1461--1476 (1998).
\item{[24]}
A. PETTERS, Morse Theory and gravitational microlensing, 
{\it J. Math. Phys.} {\bf 33}, 1915--1931 (1992).

\item{[25]}
A. PETTERS, 
Multiplane gravitational lensing I. Morse Theory and image counting,
{\it J. Math. Phys.} {\bf 36}, 4263--4275 (1995).
\item{[26]}
A. PETTERS, Lower bounds on image magnification in
gravitational lensing, 
{\it Proc. R. Soc. London A} {\bf 452}, 1475--1490 (1996).

\item{[27]}
S. REFSDAL, J.SURDEY,
Gravitational Lenses,
{\it Rep. Progr. Phys.} {\bf 56}, 117-185 (1994).
\item{[28]}
P. SCHNEIDER, J. EHLERS, E. FALCO,
{\it Gravitational lensing}.
Springer, Berlin, 1992.
\item{[29]}
J.P. SERRE, 
Homologie singuliere des espaces fibres,
{\it Ann. Math. } {\bf 54}, 425-505 (1951).
\item{[30]}
M. SPIVAK, 
{\it A comprehensive introduction to Differential Geomtery}, vol.2, 
(second edition) Publish or Perish, Inc, Wilmington, Delaware.
\item {[31]}
K. UHLENBECK,
A Morse Theory for geodesics on a Lorentz manifold,
{\it Topology} {\bf 14},  69-90 (1975).
\item {[32]}
D. WALSH, R.CARSWELL, R.WEYMANN, 
0957+561 A,B: Twin quasistellar object or gravitational lens?
{\it Nature} {\bf 279}, 381-384 (1979).
\item{[33]}
H. WEYL, 
Zur Gravitationstheorie, 
{\it Ann. Phys.} {\bf 54}, 117-145 (1917).

\end